\documentclass[fleqn,usenatbib]{mnras}

\usepackage{newtxtext,newtxmath}
\usepackage[T1]{fontenc}
\usepackage{ae,aecompl}
\usepackage{pdflscape}
\usepackage{graphicx}	
\usepackage{microtype}
\usepackage{mathrsfs}
\usepackage[flushleft]{threeparttable}

\title[AT\,2016\lowercase{dah} \& AT\,2017\lowercase{fyp}:\ novae in a tidal stream]{AT\,2016\lowercase{dah} and AT\,2017\lowercase{fyp}:\ the first classical novae discovered within a tidal stream}

\author[M.\ J.\ Darnley et al.]{M.\ J.\ Darnley,$^{1}$\thanks{E-mail: M.J.Darnley@ljmu.ac.uk}
A.\ M.\ Newsam,$^{1}$\thanks{E-mail: A.Newsam@ljmu.ac.uk}
K.\ Chinetti,$^{1,2}$
I.\ D.\ W.\ Hawkins,$^{1}$
\newauthor
A.\ L.\ Jannetta,$^{1,3}$
M.\ M.\ Kasliwal,$^{2}$
J.\ C.\ M$^\mathrm{c}$Garry,$^{1,4}$
M.\ M.\ Shara,$^{5}$
\newauthor
M.\ Sitaram,$^{1,6}$
S.\ C.\ Williams$^{7,8,9}$
\\
$^{1}$Astrophysics Research Institute, Liverpool John Moores University, IC2 Liverpool Science Park, Liverpool, L3 5RF, UK\\
$^{2}$Division of Physics, Mathematics and Astronomy, California Institute of Technology, Pasadena, CA 91125, USA\\
$^{3}$INTO Newcastle University, The INTO Building, Newcastle University, NE1 7RU, UK\\
$^{4}$Centre for Astrophysics Research, University of Hertfordshire, College Lane, Hatfield, AL10 9AB, UK\\
$^{5}$Department of Astrophysics, American Museum of Natural History, 79th Street and Central Park West, New York, NY 10024, USA\\
$^{6}$Department of Astronomy, University of Maryland, College Park, MD 20742-2421, USA\\
$^{7}$Finnish Centre for Astronomy with ESO (FINCA), Quantum, Vesilinnantie 5, University of Turku, 20014 Turku, Finland\\
$^{8}$Department of Physics and Astronomy, University of Turku, 20014 Turku, Finland\\
$^{9}$Physics Department, Lancaster University, Lancaster, LA1 4YB, UK
}

\date{Accepted 2020 April 20. Received 2020 April 20; in original form 2020 March 10}

\pubyear{2020}

\begin{document}
\label{firstpage}
\pagerange{\pageref{firstpage}--\pageref{lastpage}}
\maketitle

\begin{abstract}
AT\,2016dah and AT\,2017fyp are fairly typical Andromeda Galaxy (M\,31) classical novae. AT\,2016dah is an almost text book example of a `very fast' declining, yet uncommon, Fe\,{\sc ii}`b' (broad-lined) nova, discovered during the rise to peak optical luminosity, and decaying with a smooth broken power-law light curve. AT\,2017fyp is classed as a `fast' nova, unusually for M\,31, its early decline spectrum simultaneously shows properties of both Fe\,{\sc ii} and He/N spectral types -- a `hybrid'. Similarly, the light curve of AT\,2017fyp has a broken power-law decline but exhibits an extended flat-topped maximum. Both novae were followed in the UV and X-ray by the Neil Gehrels {\it Swift} Observatory, but no X-ray source was detected for either nova. The pair were followed photometrically and spectroscopically into their nebular phases. The progenitor systems were not visible in archival optical data, implying that the mass donors are main sequence stars. What makes AT\,2016dah and AT\,2017fyp particularly interesting is their position with respect to M\,31. The pair are close on the sky but are located far from the centre of M\,31, lying almost along the semi-minor axis of their host. Radial velocity measurements and simulations of the M\,31 nova population leads to the conclusion that both novae are members of the Andromeda Giant Stellar Stream (GSS). We find the probability of at least two M\,31 novae appearing coincident with the GSS by chance is $\sim\!1\%$. Therefore, we claim that these novae arose from the GSS progenitor, not M\,31 --- the first confirmed novae discovered in a tidal steam.
\end{abstract}


\section{Introduction}

Almost half a century ago, \citet{1972ApJ...178..623T} published a classic, monumental paper that forever changed how astronomers think about the formation and evolution of galaxies.  Their simple-titled paper ``Galactic Bridges and Tails'' demonstrated that previously unexplained, luminous connections between galaxies and long streamers emanating from those galaxies are tidal in origin. Their tour de force Figure~23 shows a model of NGC\,4038 and NGC\,4039, also known as ``The Antennae'', that mimics those galaxies' tidal tails with remarkable fidelity. The tails stretch far beyond the confines of each of the galaxies; the stars in them will never return to the galaxies in which they were born.

Rather than being arcane, evanescent features of galaxies sweeping by or through each other, tails and bridges highlight the changes in masses, sizes, morphologies and star-forming histories of galaxies that shape the appearances of the galaxies we observe today. During close galaxy--galaxy encounters, a fraction of the stars in each galaxy acquire sufficient kinetic energy to permanently escape into intergalactic space, thereby becoming ``escaped'' or hostless stars. Others travel to further than a few virial radii for longer than a few Gyr, but still remain energetically bound to their parent galaxy,  becoming ``wandering'' stars \citep*{2009ApJ...707L..22T}. The detection of these hostless and of wandering stars, and determination of their numbers and spatial distributions, is an important constraint on tidal stripping efficiency. \citet{1983ApJ...268..495M} and \citet{1984ARA&A..22..185D} stressed the importance of obtaining reliable measurements of the intracluster light as a direct indicator of the tidal damage suffered by galaxies.

Clusters of galaxies are targets amenable to searches for such intracluster stars. Hostless, or intracluster, planetary nebulae (PNe) are particularly attractive for intracluster star searches, and they have been detected, via their strong [O\,{\sc iii}] emission, in multiple clusters, including Fornax \citep{1997MNRAS.284L..11T}, Virgo (\citealt*{1998ApJ...503..109F}; \citealt{2013A&A...558A..42L}), and Coma \citep{2005ApJ...621L..93G}. Hostless type Ia supernovae have been employed as probes to indirectly measure the fraction of intracluster light \citep[see, for e.g.,][]{2003AJ....125.1087G, 2010MNRAS.403L..79M,2011ApJ...729..142S,2015ApJ...807...83G}.

Classical novae \citep[CNe; see, for e.g.,][for recent compilations of reviews]{2008clno.book.....B,2014ASPC..490.....W} are up to $100\times$ more luminous than PNe, and so sample a volume $1000\times$ larger. CNe call attention to themselves both by their 10--20 magnitude amplitude eruptions and their strong and persisting H$\alpha$ emission lines \citep[see, for e.g.,][]{1987ApJ...318..520C}. Intracluster CNe have been detected in the Fornax galaxy cluster \citep*{2005ApJ...618..692N}. 

The spatial distribution of those Fornax intracluster novae is consistent with $\sim28\pm13\%$ of the total light in the cluster being in the intracluster light \citep{2005ApJ...618..692N}. Similar fractions are evident from the Fornax PNe, while the deepest recent searches in Virgo yield an estimate of 7--15\% for the fraction of intracluster light in that cluster \citep{2017ApJ...834...16M}. This demonstrates that intracluster stars are a significant fraction of the stars in clusters, and of course their fraction will grow monotonically in the Gyr to come.

Our knowledge of hostless and wandering stars outside clusters is sparse. That they exist in the environs of the Milky Way and in the Local Group is evident from distant M-giant surveys \citep{2014AJ....147...76B}, the existence of the RR\,Lyrae and M-giant tracers in the Sagittarius Stream \citep{2001ApJ...547L.133I} and Magellanic Stream \citep{2016ARA&A..54..363D}, and the stellar streams of the Andromeda Galaxy (M\,31), particularly its Giant Stellar Stream \citep[hereafter GSS;][]{2001Natur.412...49I}. The size of the Local Group intracluster population is almost certainly smaller than that in rich clusters of galaxies, but there is currently no quantitative estimate of that size. 

The ease with which erupting CNe can be detected with 1 and 2 metre telescopes, which are now routinely carrying out automated wide-field CCD transient surveys, means that they are prime candidates for mapping hostless and wandering stars out to at least 3--5\,Mpc \citep{2006AJ....131.2980S} --- the M\,81 galaxy group and beyond. The neighbourhood of M\,31, which is being heavily targeted by several transient surveys, is a particularly useful target because the transient population in that galaxy, particularly its nova content \citep[see][for recent reviews]{2019arXiv190910497D,2019enhp.book.....S}, is extremely well-studied. 

Here we report on a pair of CNe discovered $1^\circ\!\!.2$ and $1^\circ\!\!.5$ from the centre of M\,31, close to the minor axis of their highly inclined host, and far beyond its visible optical disk. Both appear to be strongly associated with the M\,31 GSS \citep[see][and Section~\ref{sec:spatial}]{2001Natur.412...49I}. Dozens more such detections will be needed to fully map out the  hostless and wandering CNe associated with M\,31, but this Paper demonstrates that such an effort is both underway and entirely straightforward. 

AT\,2016dah (also referred to as ASASSN-16hf and iPTF\,16bqy) was discovered on 2016 July 12 by the intermediate Palomar Transient Factory \citep[iPTF;][]{2016ATel.9248....1C}, and independently two days later by the All Sky Automated Survey for Supernovae \citep[ASAS-SN;][]{2016ATel.9245....1N} survey.  The reported position of AT\,2016dah was $\alpha=0^\mathrm{h}44^\mathrm{m}41^\mathrm{s}\!.05$, $\delta=+40^\circ8^\prime35^{\prime\prime}\!\!.9$ (J2000), placing the system $1^\circ7^\prime32^{\prime\prime}$ south and $0^\circ21^\prime56^{\prime\prime}$ east of the centre of M\,31.  

\begin{figure}
\includegraphics[width=\columnwidth]{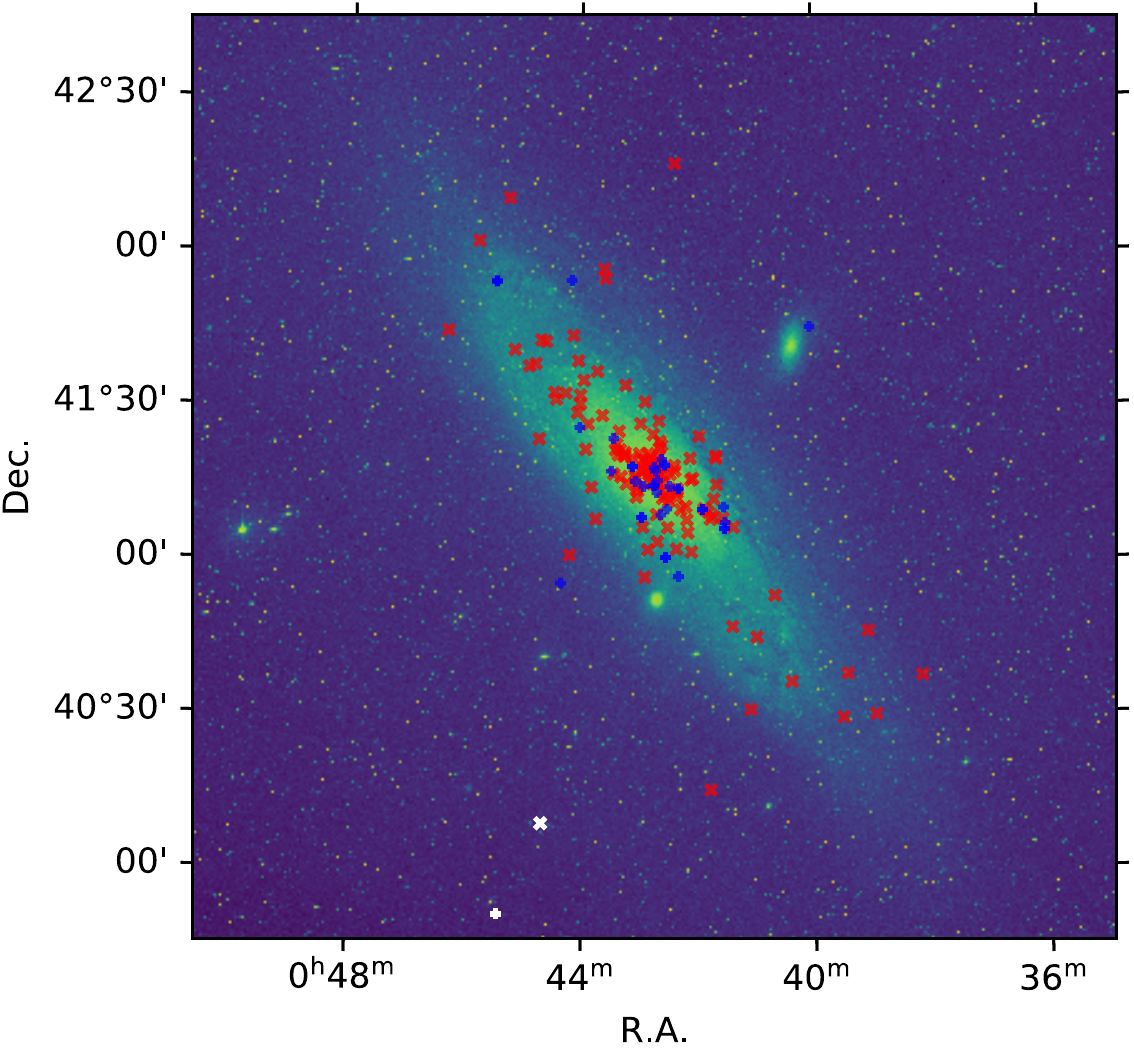}
\caption{False-colour Digitized Sky Survey \citep[DSS;][]{1990AJ.....99.2019L} red mosaic of M\,31 over-plotted with the positions of 135 spectroscopically confirmed M\,31 Fe\,{\sc ii} novae (red $\times$) and 38 He/N (including hybrids) novae (blue $+$); spectroscopic data from \citet[and C.\ Ransome et al.\ in preparation]{2011ApJ...734...12S}. The white data points show the location of AT\,2016dah (northern most; $\times$) and AT\,2017fyp (southern most; $+$).\label{M31_field}}
\end{figure}

AT\,2017fyp (aka ATLAS17jgy and Gaia17cgm) was discovered on 2017 August 7 by the Asteroid Terrestrial-impact Last Alert System \citep[ATLAS;][]{2017TNSTR.852....1T}. AT\,2017fyp is located at $\alpha=0^\mathrm{h}45^\mathrm{m}25^\mathrm{s}\!.490$, $\delta=+39^\circ50^\prime52^{\prime\prime}\!\!.34$ (J2000), $1^\circ25^\prime15^{\prime\prime}$ south and $0^\circ30^\prime11^{\prime\prime}$ east of the centre of M\,31. The positions of AT\,2016dah and AT\,2017fyp with respect to M\,31 and its spectroscopically confirmed nova population \citep[see][and C.\ Ransome et al.\ in preparation]{2011ApJ...734...12S} are illustrated in Figure~\ref{M31_field}.

In this Paper we present follow-up observations of these novae located in the outer suburbs of M\,31 and discuss the significance of their location within the M\,31 GSS. In Section~\ref{sec:observations} we describe the observations of the novae. In Section~\ref{sec:results} we go onto describe the results of the photometric, spectroscopic, and X-ray data analysis. Then in Section~\ref{sec:spatial} we explore the association of AT\,2016dah and AT\,2017fyp with the M\,31 GSS. Finally we discuss our findings in Section~\ref{sec:discussion}, before summarising our conclusions in Section~\ref{sec:conclusions}. Throughout this Paper, all quoted uncertainties are to $1\sigma$ and all lower or upper limits are evaluated at $3\sigma$, unless otherwise stated.

\section{Observations}\label{sec:observations}	

\subsection{AT 2016dah}

Photometric observations of AT\,2016dah were obtained using the 2.0\,m fully robotic Liverpool Telescope \citep[LT;][]{2004SPIE.5489..679S}, La Palma, and the $48^{\prime\prime}$ Samuel Oschin telescope (P48) at Palomar.  LT imaging was taken using the IO:O CCD camera\footnote{\url{http://telescope.livjm.ac.uk/TelInst/Inst/IOO}} \citep{ljmu5699} through $u'BVr'i'$ filters, while P48 images were obtained using the CFH12K CCD mosaic camera through an $r'$ filter.  Additional photometric data were also obtained via the All-Sky Automated Survey for Supernovae \citep[ASAS-SN;][]{2014ApJ...788...48S,2017PASP..129j4502K} Sky Patrol\footnote{\url{https://asas-sn.osu.edu}}.

The LT photometric data were reduced using tools within the IRAF environment \citep{1993ASPC...52..173T}, and aperture photometry was performed using the {\tt qphot} task. The LT data were calibrated against 81 stars in the field using photometry from Pan-STARRS \citep[DR1;][]{2016arXiv161205560C}, these sources each had Pan-STARRS $griz$ magnitudes in the range $14\!<\!m\!<\!19$ (with catalogue uncertainties $\Delta m\!\leq\!0.1$). The $u'$-band data were calibrated using a subset of 34 of those field stars that contained photometry in the range $14\!<\!u'\!<\!19$ ($\Delta u'\!\leq\!0.1$) from data release \#12 of the Sloan Digital Sky Survey \citep[SDSS;][]{2015ApJS..219...12A}.  The photometry of the standards was converted from Sloan to Johnson magnitudes, where required, using the appropriate transformations from \citet{2005AJ....130..873J}.  The subsequent LT photometry is reproduced in Table~\ref{16full_photometry} and information about the secondary standards is presented in Table~\ref{16full_standards}. A full overview of the photometric process employed is given in \citet{2007ApJ...661L..45D,2016ApJ...833..149D}. The P48 (iPTF) data were obtained directly from that survey's near real-time discovery pipeline (\citealt*{2016PASP..128k4502C}; \citealt{2017PASP..129a4002M}).

Spectroscopic observations were obtained using the SPRAT \citep{2014SPIE.9147E..8HP} instrument on the LT operating in the blue-optimised mode.  A slit width of $1^{\prime\prime}\!\!.8$ was used, resulting in a spectral resolution of $\sim$20\,\AA, or a velocity resolution of $\sim$1000\,km\,s$^{-1}$. A log of the spectroscopic observations is provided in Table~\ref{spec_log}.

\begin{table}
\caption{Summary of all spectroscopic observations of AT\,2016dah and AT\,2017fyp with the SPRAT spectrograph on the Liverpool Telescope.\label{spec_log}}
\begin{center}
\begin{tabular}{lll}
\hline
Date [UT] & $\Delta t$ [days] & Exposure [s]  \\
\hline
\multicolumn{3}{c}{AT\,2016dah}\\
\hline
2016 Jul 14.084 & \phantom{0}2.124 & $2\times600$ \\
2016 Jul 14.135 & \phantom{0}2.175 & $3\times600$ \\
2016 Jul 15.077 & \phantom{0}3.117 & $3\times600$ \\
2016 Jul 16.107 & \phantom{0}4.174 & $3\times600$ \\
2016 Jul 19.101 & \phantom{0}7.141 & $3\times600$ \\
2016 Jul 26.135 & 14.115 & $3\times600$ \\
2016 Aug 03.093 & 22.133 & $3\times600$ \\
2016 Aug 05.109 & 24.149 & $3\times600$ \\
2016 Aug 09.094 & 28.134 & $2\times600$ \\
2016 Aug 13.113 & 32.153 & $3\times600$ \\
2016 Aug 22.093 & 41.133 & $3\times600$ \\
2016 Aug 29.050 & 48.090 & $5\times600$ \\
2016 Sep 13.127 & 63.167 & $2\times600$ \\
\hline
\multicolumn{3}{c}{AT\,2017fyp}\\
\hline
2017 Aug 11.097 & \phantom{0}5.017 & $3\times600$ \\
2017 Aug 13.018 & \phantom{0}6.938 & $3\times600$ \\
2017 Aug 15.081 & \phantom{0}9.001 & $3\times600$ \\
2017 Aug 17.132 & 11.052 & $3\times600$ \\
2017 Aug 20.117 & 14.037 & $3\times600$ \\
2017 Aug 26.105 & 20.025 & $3\times600$ \\
2017 Sep 01.040$^\dag$ & 25.960 & $3\times600$ \\
2017 Sep 16.045 & 40.965 & $3\times900$ \\
2017 Sep 30.126 & 55.046 & $3\times900$ \\
2017 Oct 18.939 & 73.859 & $3\times900$ \\
2017 Nov 12.912 & 98.832 & $3\times 900$ \\
\hline
\end{tabular}
\end{center}
\begin{tablenotes}
\small
\item $^\dag$ The night of 2017 Aug 31 was non-photometric and this spectrum was collected through cloud. We have not included it in the analysis due to the poor signal-to-noise.
\end{tablenotes}
\end{table}

Initial reduction of the LT spectra, which includes bias subtraction, flat field correction, and sky subtraction, up to point of wavelength calibration, was performed using the SPRAT data reduction pipeline (see \citealt*{2012AN....333..101B}; \citealt{2014SPIE.9147E..8HP}). To perform relative flux calibration, we utilised 77 observations of the spectrophotometric standard star Hilt\,102 \citep{1977ApJ...218..767S}, taken on photometric nights between 2017 June 19 and 2017 Dec 31\footnote{Prior to these dates, standards were not routinely taken by SPRAT.}, to construct a master sensitivity function. As the sensitivity function was constructed from data collected almost a year post-eruption, the absolute flux calibration was modified by comparison between bandpass spectrophotometry and the LT or iPTF $r'$-band data. As such we estimate that the flux uncertainty of the AT\,2016dah data is between 10--15\%.

A Neil Gehrels {\it Swift} Observatory \citep{2004ApJ...611.1005G} target of opportunity (ToO) request was approved shortly after spectroscopic confirmation of AT\,2016dah (Target ID:\ 34619).  Beginning 6.6\,d post-discovery, eleven approximately weekly {\it Swift} visits were utilised each with a target exposure time of 2\,ks.  The {\it Swift} UV/optical telescope \citep[UVOT;][]{2005SSRv..120...95R} was employed using the uvw1 filter ($\sim2600$\,\AA).  The {\it Swift} X-ray Telescope \citep[XRT;][]{2005SSRv..120..165B} was also deployed, in the photon counting (PC) mode. 

The Swift data were processed using the HEASoft (v6.26.1 for UVOT; v6.22 for XRT) software and the corresponding calibration files. UVOT magnitudes were calculated using the {\tt uvotsource} tool, with a standard 5$^{\prime\prime}$ radius circular extraction region for the source, and a 46$^{\prime\prime}$ radius circular aperture offset from, but close to ($\alpha=0^\mathrm{h}44^\mathrm{m}43^\mathrm{s}$, $\delta=40^\circ09^\prime34^{\prime\prime}$), the source used to estimate the background. The UV emission had faded below detectability by the end of September. The XRT data were analysed using the freely available on-line tool\footnote{\url{https://www.swift.ac.uk/user_objects}} \citep{2009MNRAS.397.1177E}. No X-ray source corresponding to AT\,2016dah was detected at any time \cite[also see][]{2016ATel.9329....1C}. Results from the UVOT and XRT analysis are presented in Table~\ref{swift_data}.

\begin{table*}
\caption{Neil Gehrels {\it Swift} Observatory XRT and UVOT observations of AT\,2016dah and AT\,2017fyp.\label{swift_data}}
\begin{center}
\begin{tabular}{lllllllll}
\hline
Date & $\Delta t$ & \multicolumn{2}{c}{MJD $57000+$} & Exposure & \multicolumn{2}{c}{XRT (0.3--10\,keV)} & UVOT uvw1 & Obs.\ ID  \\
\cline{3-4} \cline{6-7}
[UT] & [days] & Start & End & time$^\mathrm{a}$\,[s] & [counts\,sec$^{-1}$] & $L$\,[$10^{36}$\,erg\,s$^{-1}$] & [mag]$^\mathrm{b}$\\
\hline
\multicolumn{9}{c}{AT\,2016dah}\\
\hline
2016 Jul 19.377 & \phantom{0}7.417 & \phantom{0}588.008 & \phantom{0}588.746 & 2041.8 & $<0.008$ & $<\phantom{0}3.6$ & $17.12\pm0.03$ & 00034619002\\
2016 Jul 26.278 & 14.318 & \phantom{0}595.237 & \phantom{0}595.319 & 2044.1& $<0.005$ & $<\phantom{0}2.3$ & $18.00\pm0.05$ & 00034619003\\
2016 Aug 03.621 & 22.661 & \phantom{0}603.356 & \phantom{0}603.887 & 1183.4& $<0.032$ & $<14.4$ & $18.25\pm0.07$ & 00034619004\\
2016 Aug 11.903 & 30.943 & \phantom{0}611.864 & \phantom{0}611.941 & 1847.0& $<0.006$ & $<\phantom{0}2.7$ &  $19.11\pm0.09$ & 00034619005\\
2016 Aug 20.677 & 39.717 & \phantom{0}620.642 & \phantom{0}620.712 & 1181.8 & $<0.007$ & $<\phantom{0}3.2$ & $19.36\pm0.13$ & 00034619006\\
2016 Sep 04.695 & 54.735 & \phantom{0}635.654 & \phantom{0}635.736 & 2066.8& $<0.004$ & $<\phantom{0}1.8$ &  $20.26\pm0.19$ & 00034619007\\
2016 Sep 12.799 & 62.839 & \phantom{0}643.768 & \phantom{0}643.830 & 1281.0& $<0.019$ & $<\phantom{0}8.5$ &  $20.49\pm0.29$& 00034619008\\
2016 Sep 20.773 & 70.813 & \phantom{0}651.732 & \phantom{0}651.814 & 1902.2& $<0.007$ & $<\phantom{0}3.2$ &  $20.50\pm0.23$& 00034619009\\
2016 Sep 29.772 & 79.812 & \phantom{0}660.767 & \phantom{0}660.778 & \phantom{0}954.5& $<0.044$ & $<19.8$ & \ldots$^\mathrm{c}$ & 00034619011\\
2016 Oct 02.889 & 82.929 & \phantom{0}663.888 & \phantom{0}663.889 & \phantom{0}116.3& $<0.157$ & $<70.6$ & $>19.2$ & 00034619012\\
2016 Oct 06.714 & 86.754 & \phantom{0}667.675 & \phantom{0}667.752 & 1966.5& $<0.005$ & $<\phantom{0}2.3$ & ($21.09\pm0.38$)$^\mathrm{d}$ & 00034619013\\
\hline
\multicolumn{9}{c}{AT\,2017fyp}\\
\hline
2017 Aug 24.802 & \phantom{0}18.722 & \phantom{0}989.702 & \phantom{0}989.901 & 1836.0 & $<0.015$ & $<\phantom{0}6.2$ & $16.40\pm0.03$ & 00010239001\\
2017 Sep 07.101 & \phantom{0}32.021 & 1003.002 & 1003.200 & \phantom{0}451.2 & $<0.021$ & $<\phantom{0}8.7$ & $17.14\pm0.06$ & 00010239002\\
2017 Sep 14.504 & \phantom{0}39.424 & 1010.368 & 1010.641 & 1401.0 & $<0.020$ & $<\phantom{0}8.2$ & $17.73\pm0.05$ & 00010239003\\
2017 Oct 18.279 & \phantom{0}73.199 & 1044.116 & 1044.441 & 3695.4 & $<0.003$ & $<\phantom{0}1.2$ & $18.65\pm0.05$ & 00010239004\\
2017 Nov 18.190 & 104.110 & 1075.115 & 1075.264 & 4049.2 & $<0.002$ & $<\phantom{0}0.8$ & $19.36\pm0.08$ & 00010239005\\
2017 Dec 18.593 & 134.513 & 1105.224 & 1105.962 & 3149.9 & $<0.004$ & $<\phantom{0}1.6$& $20.43\pm0.17$ & 00010239006\\
2018 Jan 18.776 & 165.696 & 1136.773 & 1136.778 & \phantom{0}425.9 & $<0.051$ & $<21.0$ & $>20.0$ & 00010239007\\
2018 Jan 20.492 & 167.412 & 1138.490 & 1138.494 & \phantom{0}322.7 & $<0.025$ & $<10.3$ & $>19.6$ & 00010239008\\
2018 Jan 21.961 & 168.881 & 1139.760 & 1140.162 & 1153.8 & $<0.011$ & $<\phantom{0}4.5$ & $>20.5$ & 00010239009\\
2018 Jan 23.920 & 170.840 & 1141.887 & 1141.953 & 2209.9 & $<0.006$ & $<\phantom{0}2.5$ & $>20.9$ & 00010239010\\
\hline
2018 Jan 18--24$^\mathrm{e}$ & 168.284 & 1136.773 & 1141.953 & 3822.3 & $<0.003$ & $<\phantom{0}1.2$ & $>21.2$ & \ldots007---010 \\
\hline
\end{tabular}
\end{center}
\begin{tablenotes}
\small
\item $^\mathrm{a}$Here the exposure time refers specifically to UVOT.
\item $^\mathrm{b}$UVOT magnitudes are reported in the Vega system, we quote the random statistical uncertainties, calibration systematic uncertainties are 0.03\,mag.
\item $^\mathrm{c}$No useable data due to loss of star tracker lock.
\item $^\mathrm{d}$Source detection significance at the nova position was at $2.9\sigma$.
\item $^\mathrm{e}$Here we combine the final four {\it Swift} observations as they were taken within a one week period.
\end{tablenotes}
\end{table*}

For each {\it Swift} XRT observation, we have also estimated the upper luminosity limit of the source (0.3--10\,keV). Here we assumed a distance to M\,31 of $d=752\pm17$\,kpc \citep{2001ApJ...553...47F}, an estimate of the Galactic neutral atomic H density $n_\mathrm{H}=5.32\times10^{20}\,\mathrm{cm}^{-2}$ \citep{2016A&A...594A.116H}, and a typical (for X-ray detected M\,31 novae) blackbody temperature of $kT=50$\,eV \citep{2014A&A...563A...2H}. Luminosity limits were estimated using the web interface to PIMMS (v4.10b)\footnote{\url{https://heasarc.gsfc.nasa.gov/cgi-bin/Tools/w3pimms/w3pimms.pl}} and are recorded in Table~\ref{swift_data}.

\subsection{AT 2017fyp} 

The LT photometric data for AT\,2017fyp were reduced in a similar manner as those for AT\,2016dah. Here, the LT data were calibrated against 42 stars in the field using photometry from SDSS DR12, these sources each had Sloan $u'g'r'i'z'$ magnitudes in the range $14\!<\!m\!<\!19$ ($\Delta m\!\leq\!0.1$). The subsequent LT photometry is reproduced in Table~\ref{17full_photometry} and information about the secondary standards is presented in Table~\ref{17full_standards}.

The Las Cumbres Observatory \citep[LCO;][]{2013PASP..125.1031B} 2.0\,m telescope at Haleakala (formally the Faulkes Telescope North) was used to collect four additional epochs of $BVr'i'$ photometry. These data were taken using the Spectral camera\footnote{\url{https://lco.global/observatory/instruments/spectral}} and were reduced and analysed in an identical manner to the LT imaging data. The LCO photometry is also recorded within Table~\ref{17full_photometry}.

Additional photometric data for AT\,2017fyp, limited to the pre-discovery and discovery photometry, were obtained from the ATLAS survey \citep{2017TNSTR.852....1T}. Three {\it Gaia} photometry points are also available\footnote{\url{http://gsaweb.ast.cam.ac.uk/alerts/alert/Gaia17cgm}}.

The LT SPRAT spectroscopic data were reduced in an identical manner to those for AT\,2016dah, and we employed the same sensitivity function to produce relative flux calibrated spectra. However, for these data, no discernible improvement in the absolute flux calibration was found by using bandpass spectrophotometry, in part this was due to the lower photometric cadence. We estimate the flux uncertainty of the AT\,2017fyp data is between 15--20\%.

A {\it Swift} ToO programme was also rapidly approved post-discovery for AT\,2017fyp (Target ID:\ 10239) and a series of ten observations commenced two weeks post-discovery. These {\it Swift} XRT and UVOT data were reduced in an identical fashion to those for AT\,2016dah. For the UVOT photometry, a circular background estimation region of radius 67$^{\prime\prime}$ was positioned at $\alpha=0^\mathrm{h}45^\mathrm{m}34^\mathrm{s}$, $\delta=39^\circ49^\prime25^{\prime\prime}$. Like AT\,2016dah, no X-ray source was detected, here upper luminosity limits were derived assuming an average Galactic column of $n_\mathrm{H}=4.73\times10^{20}\,\mathrm{cm}^{-2}$. The AT\,2017fyp swift results are also reported in Table~\ref{swift_data}. The final four {\it Swift} visits were obtained within a single week, combining them does not yield either an XRT or UVOT detection at that time.

\subsection{Archival data}\label{archives}

To enable a search for the progenitor systems of AT\,2016dah and AT\,2017fyp, i.e.\ the quiescent novae, we utilised a number of archival observations. The positions of both AT\,2016dah and AT\,2017fyp are located within Field 7 of the Canada-France-Hawaii Telescope (CFHT) survey of the M\,31 GSS \citet[also see Sections~\ref{sec:spatial} and \ref{sec:stream}]{2003MNRAS.343.1335M}. These data were taken using the CFH12K instrument, a $12,288\times8,192$ pixel CCD mosaic camera \citep{2000SPIE.4008.1010C} and consist of a pair of $3\times545$\,s exposures through Mould $V$ and $I$-band filters. The CFHT data were collected on 2001 September 13, $\sim15$\,yrs before the eruptions. AT\,2016dah lies within chip02 of Field 7, whereas AT\,2017fyp can be found within chip09. As the field of view of each of the CFH12K chips is roughly similar to the LT field, these data were processed in a similar fashion to the LT data, including stacking of each set of three images, and photometry was performed as described above. We also utilised the Galaxy Evolution Explorer \citep[GALEX;][]{2005ApJ...619L...1M} data archive to extend the search to the near- and far-UV.

A search for potential progenitor/quiescent systems was performed following the procedure set out in \citet{2009ApJ...705.1056B}, and then further refined by \citet{2014ApJS..213...10W} and more recently \citet{2019MNRAS.486.4334H}. Astrometric solutions were computed between each $V$ and $i'$-band LT observation and the corresponding CFHT $V$ and $I$-band stack, and the GALEX data, to define the progenitor search region with those data.

\section{Results}\label{sec:results}

\subsection{Time of eruption}

Although the last pre-discovery observation of AT\,2016dah was taken in the $V$-band by ASAS-SN on 2016 Jul 11.52\,UT, this observation was not of sufficient depth to rule out an eruption. Therefore, we assume that the eruption occurred between the last iPTF non-detection ($R>21.171$) on Jul 11.48 and the iPTF detection ($R=18.78\pm0.07$) on Jul 12.44. For the purpose of our analysis, we take the time of eruption to be 2016 Jul $11.96\pm0.48$\,UT (MJD:\ $57580.96\pm0.48$). 

AT\,2017fyp was observed by ATLAS on 2017 Aug 04.603\,UT; no source was detected down to a limit of 19.28\,mag (through the ATLAS \textit{orange} filter). We again assume that the eruption occurred at some point between that observation and the discovery observation, also by ATLAS, on Aug 07.553. Therefore we take the time of eruption of AT\,2017fyp to be 2017 Aug $6.08\pm1.48$\,UT (MJD:\ $57971.08\pm1.48$).

\subsection{Reddening}

The general Galactic reddening toward M\,31 was determined as $E\left(B-V\right)=0.1$ \citep{1992ApJS...79...77S}. \citet*{1998ApJ...500..525S} find $E\left(B-V\right)=0.06$ and 0.05 around the positions of AT\,2016dah and AT\,2017fyp, respectively. Converting the Galactic $n_\mathrm{H}$ estimates (see Section~\ref{sec:observations}) to reddening \citep[using Equation~1 from][and assuming $R_V$=3.1]{2009MNRAS.400.2050G} we find $E\left(B-V\right)=0.08$ and 0.07 for AT\,2016dah and AT\,2017fyp, respectively. There are no reliable reddening markers available for nova photometry, and there are none present in our obtained spectra with which we could independently constrain the reddening toward either system. Some authors have used the Balmer decrement to estimate the reddening for particular novae, however others \citep*[see, for e.g., for a recent discussion][]{2017MNRAS.472.1300W} have shown this to be unreliable. As such, we will assume a reddening of $E\left(B-V\right)=0.1$ toward both novae\footnote{The recurrent nova M31N\,2008-12a lies in front of the bulk dust and gas in the M\,31 structural model proposed by \citet{2015ApJ...814....3D}. \citet{2017ApJ...847...35D} directly measured the reddening toward M31N\,2008-12a to be $E\left(B-V\right)=0.10\pm0.03$, which is consistent with the \citet{1992ApJS...79...77S}, \citet{1998ApJ...500..525S}, and \citet{2016A&A...594A.116H} determinations.}. Given that both novae are located far from the M31 disk, we assume that there is negligible extinction contribution beyond the confines of the Milky Way. All spectra have been dereddened using this value, and all subsequent discussion of the spectra refer to analysis of the dereddened observations. 

\subsection{Photometric evolution}

\subsubsection{AT\,2016dah}\label{dah_phot}

\begin{figure*}
\includegraphics[width=\textwidth]{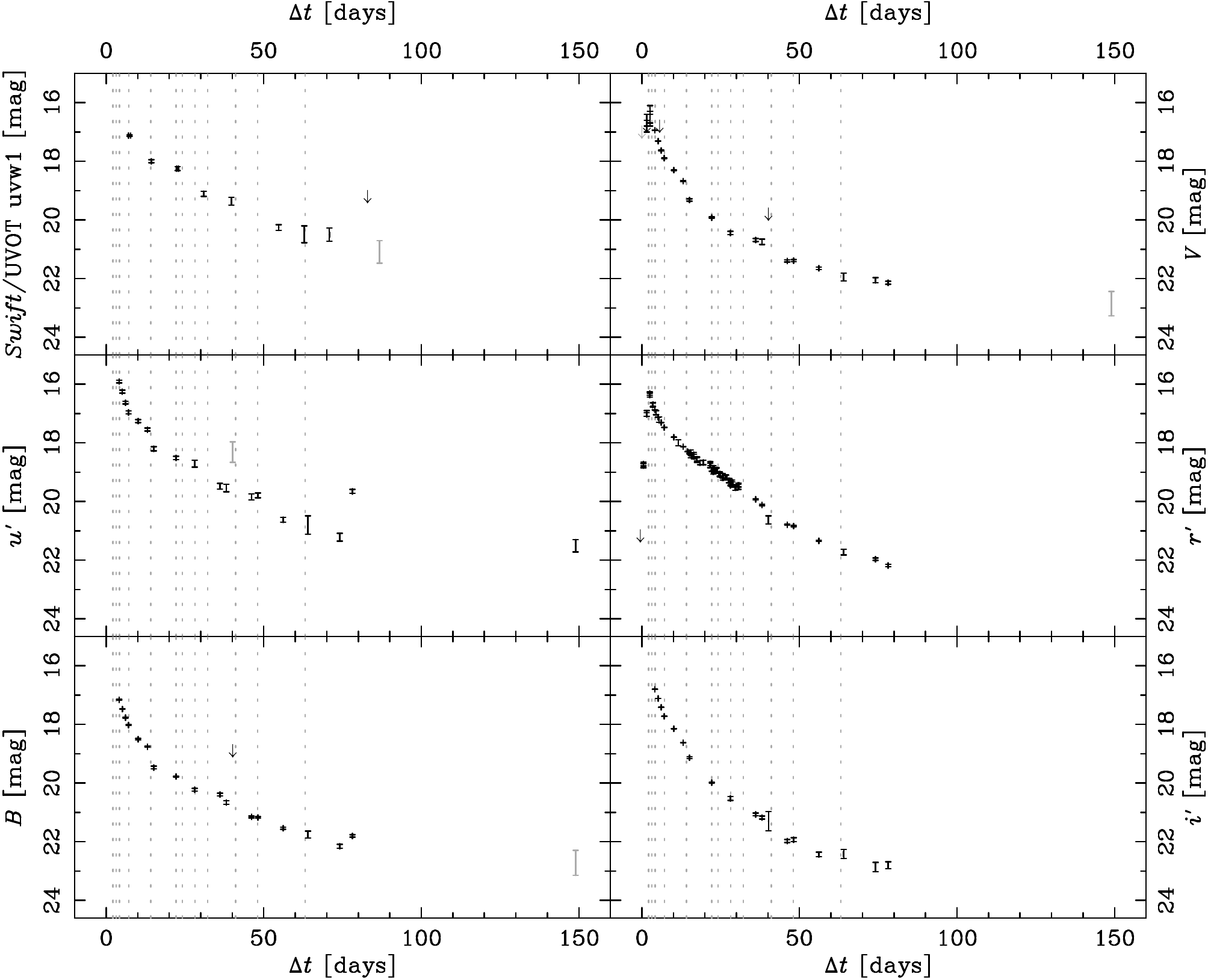}
\caption{The near-UV---optical light curve of nova AT\,2016dah.  The epochs of spectroscopy are also indicated by the vertical grey-dotted lines. Grey data points indicate detections with significance $2<\sigma\le3$.\label{dah_lc}}
\end{figure*}

The combined ASAS-SN, iPTF, LT, and {\it Swift} UVOT light curve of the eruption of AT\,2016dah are shown in Figure~\ref{dah_lc}. These light curves reveal a nova that was detected on the rise (still relatively uncommon for extragalactic novae, particularly those beyond the Magellanic Clouds), that in general follows a rapid and smooth decline through around six magnitudes (before it became undetectable). The \citet*{2010AJ....140...34S} light curve scheme would classify AT\,2016dah as a smooth or ``S''-type nova. 

\begin{figure*}
\includegraphics[width=\columnwidth]{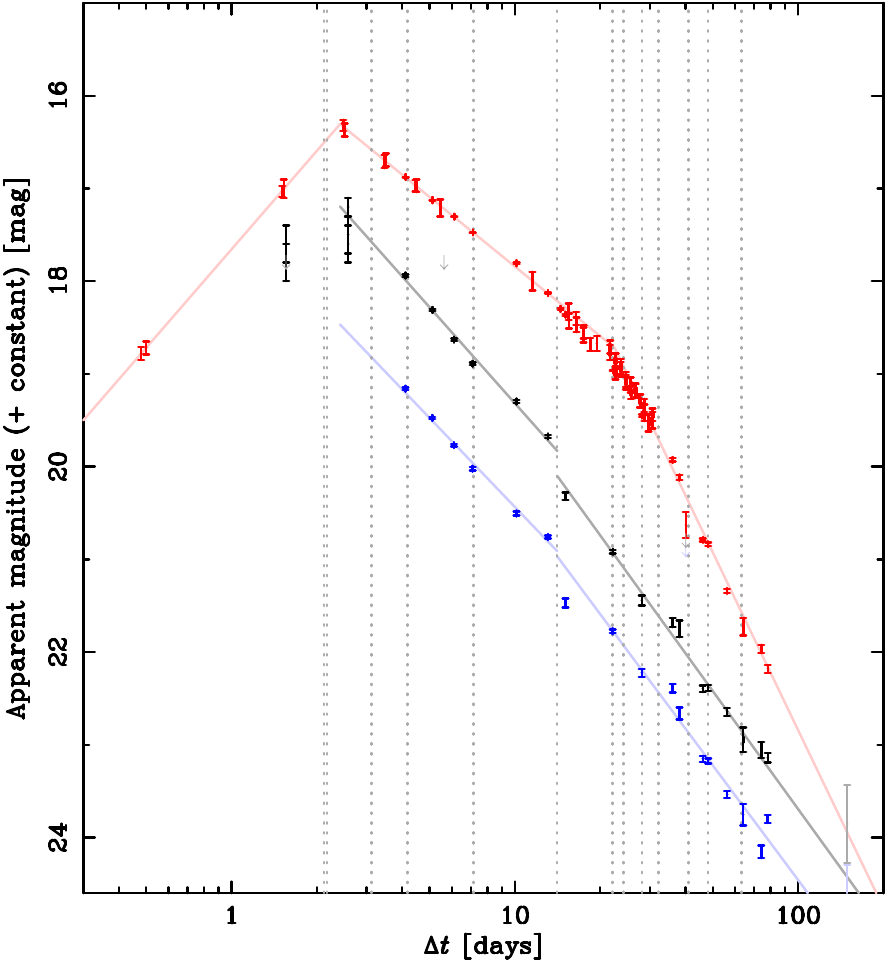}\hfill
\includegraphics[width=\columnwidth]{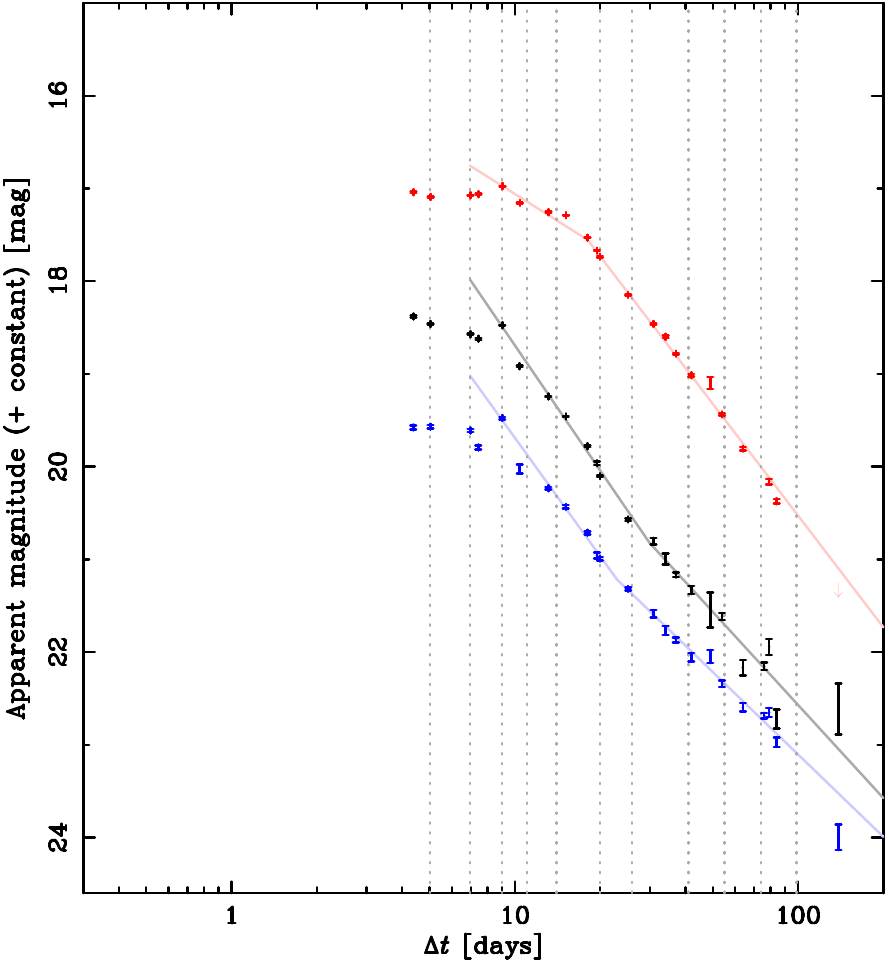}
\caption{Light curve of AT\,2016dah (left) and AT\,2017fyp (right) displayed with logarithmic time axes. The $r'$-band light curve is shown in red (top), $V$-band in black (middle; all magnitudes shifted by $+1$), an $B$-band in blue (bottom; shifted by $+2$). The broken power-law fits to these data ($F\propto t^{\alpha}$) are discussed in the text. The vertical dotted lines indicate spectroscopic epochs.\label{dah_lc_log}}
\end{figure*}

In Figure~\ref{dah_lc_log} we present the $r'$-band light-curve (red) of AT\,2016dah, plotted on a logarithmic time axis. The three straight-line fits to these data, demonstrate that the $r'$-band evolution follows three broken power-laws (between flux and time; $F\propto t^{\alpha}$). The post-maximum break occurs at around $22.5$\,d post-eruption, before the break the slope is $\alpha=-1.00\pm0.01$; post-break, $\alpha=-2.55\pm0.06$. We note that neither value is consistent with those determined by \citep{2006ApJS..167...59H}. The slope of the $r'$-band light cure before maximum light is $\alpha=1.41\pm0.01$. 

Figure~\ref{dah_lc_log} also shows the $V$-band (black) and $B$-band (blue) light-curves. While much less densely sampled than the $r'$-band data, these light curves follow a similar form, with one marked difference. There is an apparent `discontinuity' in the light curves at around day 14 (a spectrum was also taken at this time, see Section~\ref{sec:dah_spec}), where there is a `sudden' drop in the brightness by just over half a magnitude. This drop appears also appears in the $u'$-band and there is a non-conclusive hint of such an occurrence in the {\it Swift} UVOT data, but is not present in the two reddest filters. The $V$ and $B$-band data from maximum light until this drop are consistent with power-law of indices of $\alpha_V=-1.38\pm0.06$ and $\alpha_B=-1.30\pm0.05$, respectively, i.e.\ they are consistent with each other, but not with the $r'$-band decline during this phase. Post-drop, the $V$ and $B$-band declines remain consistent with indices of  $\alpha_V=-1.69\pm0.05$ and $\alpha_B=-1.65\pm0.07$, respectively. Both these later time $V$ and $B$-band declines are consistent with the power-law slopes expected during this period between $\sim2$ to $\sim6$ magnitudes below peak \citep[$\alpha=-1.75$;][]{2006ApJS..167...59H}.

For the purpose of this analysis, we will assume that maximum light occurred at the time of the brightest $r'$-band observation, $r'=16.32\pm0.06$, 2.48\,d post-eruption. A pair of spectra taken $0.33$\,d earlier (see Section~\ref{sec:dah_spec}) are consistent with pre-maximum evolution, a spectrum taken at $\Delta t=3.117$ shows the nova in a post-maximum state. As such, we can use the power-law fits to derive the following estimates of the $t_2$ and $t_3$ decline times (the time taken to decay by two and three magnitudes, respectively, from peak):\ $t_2\left(r'\right)=13.3^{+0.6}_{-0.3}$\,d (a decay rate of $\sim0.15$\,mag\,d$^{-1}$); $t_3\left(r'\right)=26\pm2$\,d and occurs $\sim6$\,d post-break. Such decline times would class AT\,2016dah as a fast nova \citep{1964gano.book.....P}. For the $V$ and $B$-bands we find, $t_2\left(V\right)=7\pm1$\,d, $t_2\left(B\right)=8\pm1$\,d, $t_3\left(V\right)=13\pm1$\,d, $t_3\left(B\right)=16^{+3}_{-2}$\,d. These values are smaller than those computed using the $r'$-band data, whose decline is slowed by the persistence of the H$\alpha$ emission line. The $V$ and $B$-band values classify AT\,2016dah as a {\it very} fast nova, the $V$-band value for $t_3$ is consistent with the epoch of the light curve `drop', the uncertainty on the $B$-band $t_3$ is larger due to the lack of observations at maximum light, but it is formally also consistent with the drop.

The light curve of AT\,2016dah was followed for around 150 days post-eruption, at which point the nova was detected $\sim6$ magnitudes below peak in the $u'$-band, and was marginally detected (sub $3\sigma$) with a similar amplitude in the $V$ and $B$-bands.

\subsubsection{AT\,2017fyp}\label{fyp_phot}

\begin{figure*}
\includegraphics[width=\textwidth]{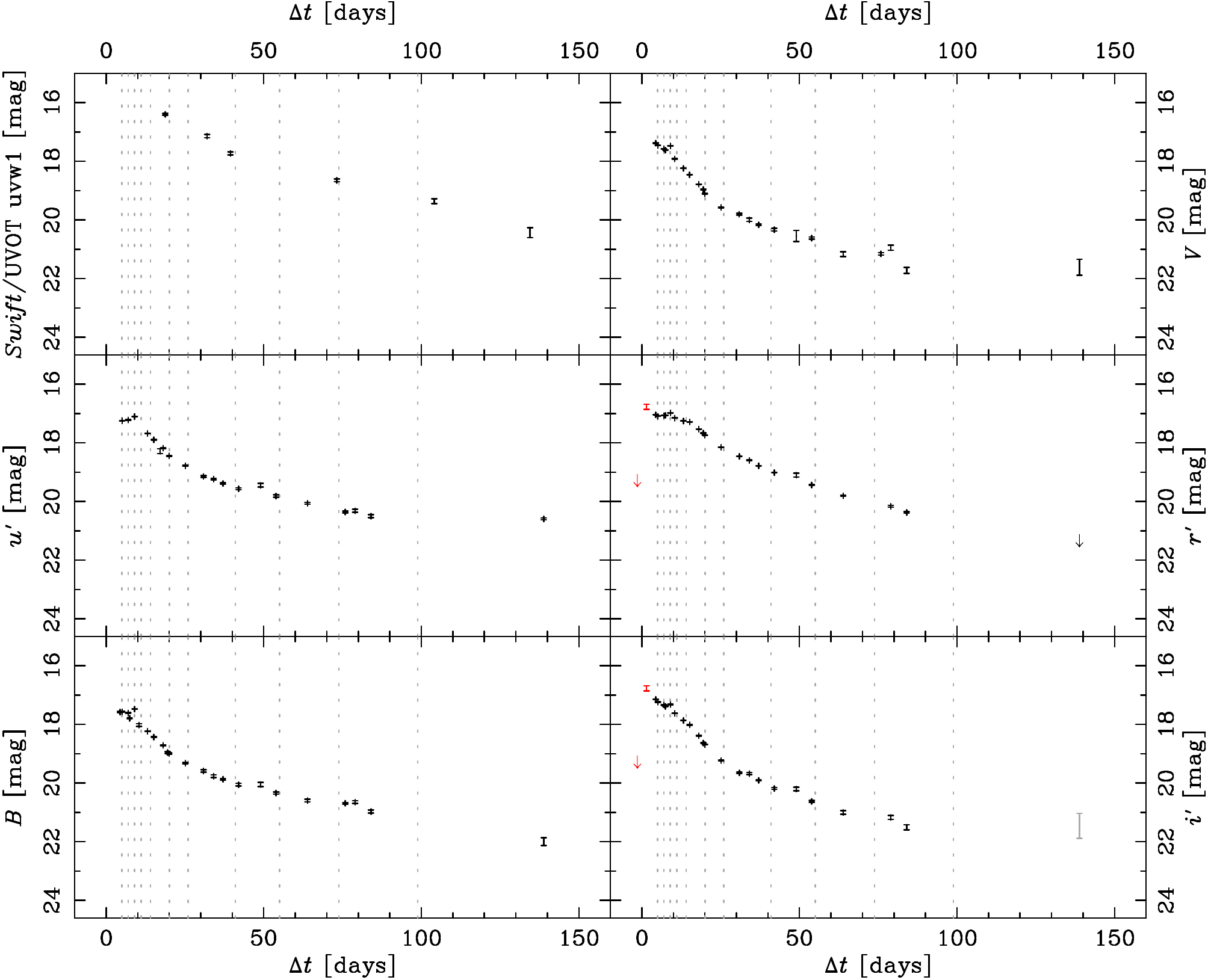}
\caption{The near-UV---optical light curve of nova AT\,2017fyp.  The epochs of spectroscopy are also indicated by the vertical grey-dotted lines. Grey data points indicate detections with significance $2<\sigma\le3$. The red data points included in the $r'$ and $i'$ indicate the ATLAS \textit{orange} filter pre-discovery upper-limit and discovery photometry; the \textit{orange} filter has a wide bandpass that covers $r'+i'$. The final four {\it Swift} data points ($\Delta t>50$\,d; see Table~\ref{swift_data}) are not shown, all four are upper limits consistent with the expected late-decline of the eruption. \label{fyp_lc}}
\end{figure*}

In Figure~\ref{fyp_lc} we present the ATLAS, LCO, LT, and {\it Swift}/UVOT photometric light curves of AT\,2017fyp. At first inspection, the evolution of AT\,2017fyp is clearly slower than that of AT\,2016dah, and it is around $\sim1$\,mag fainter at peak. There is no coverage of the rise, and maximum light may have been missed, but the light curve appears to have a short flat-top peak lasting around $\sim5$\,d. Following this flat period, the nova enters a rapid decline until a clear break in the light-curve at $\sim25$\,d post-eruption where the decline slows.  AT\,2017fyp was followed for $\sim140$\,d post-eruption through a decay of 4--5\,magnitudes, after which it remained detected in the bluer bands (and was marginally detected, $<3\sigma$, in the $i'$-band). Given the flat topped nature, and power-law like decline (see below), AT\,2017fyp may be best classed as a flat topped or ``F''-type nova. A slight increase in the luminosity of the nova during this maximum light  `plateau' phase is evident, particularly to the blue. 

As with AT\,2016dah, we present the $BVr'$-band light curves of AT\,2017fyp in Figure~\ref{dah_lc_log}, where they are plotted against a logarithmic time axis. Here the flat topped nature is evident, as is the broken-power law decline, with that break in the opposite direction in the $V$ and $B$-bands to that of AT\,2016dah (steep to shallow, for AT\,2017fyp, versus shallow to steep for AT\,2016dah). Here the $r'$-band behaviour is again probably driven by the H$\alpha$ emission line evolution.

The initial indices of the $BVr'$-band power-law declines are $\alpha=-1.69\pm0.07$, $-1.79\pm0.06$, and $-0.77\pm0.08$, respectively. Here the $V$ and $B$-band slopes are consistent with those determined by \citet{2006ApJS..167...59H}, but over an earlier part of the light curve development (as measured by decline from peak) as suggested by those authors. The later portion of the decline is consistent with power-law indices of  $\alpha=-1.18\pm0.04$, $-1.35\pm0.05$, and $-1.60\pm0.04$, for the $BVr'$-bands, respectively. 

With novae with flat topped or cusp-like (``C''-type) behaviour around peak, it is always a bone of contention how one should determine the epoch of maximum light, and hence how to reliably estimate the decline times. For this analysis, we will assume that maximum light occurred at some point during the $\sim5$\,d flat top (but no later than $\Delta t\approx9$\,d post-eruption), and that the decline time range is determined by the power-law decline and the width of the flat top. As such, we estimate the following decline times:\ 
$32\lesssim t_2\left(r'\right)\lesssim37$\,d, 
$16\lesssim t_2\left(V\right)\lesssim21$\,d, 
$20\lesssim t_2\left(B\right)\lesssim25$\,d, 
$63\lesssim t_3\left(r'\right)\lesssim68$\,d, 
$38\lesssim t_3\left(V\right)\lesssim43$\,d, 
and $53\lesssim t_3\left(B\right)\lesssim58$\,d.
The $V$ and $B$-band decline times would class AT\,2017fyp as a fast nova, whereas the slower $r'$-band times would class this eruption as {\it moderately} fast, again the influence of the H$\alpha$ emission line will have impacted the $r'$-band estimates. 

\subsection{Spectroscopic evolution}

\subsubsection{AT\,2016dah}\label{sec:dah_spec}

\begin{figure*}
\includegraphics[width=\textwidth]{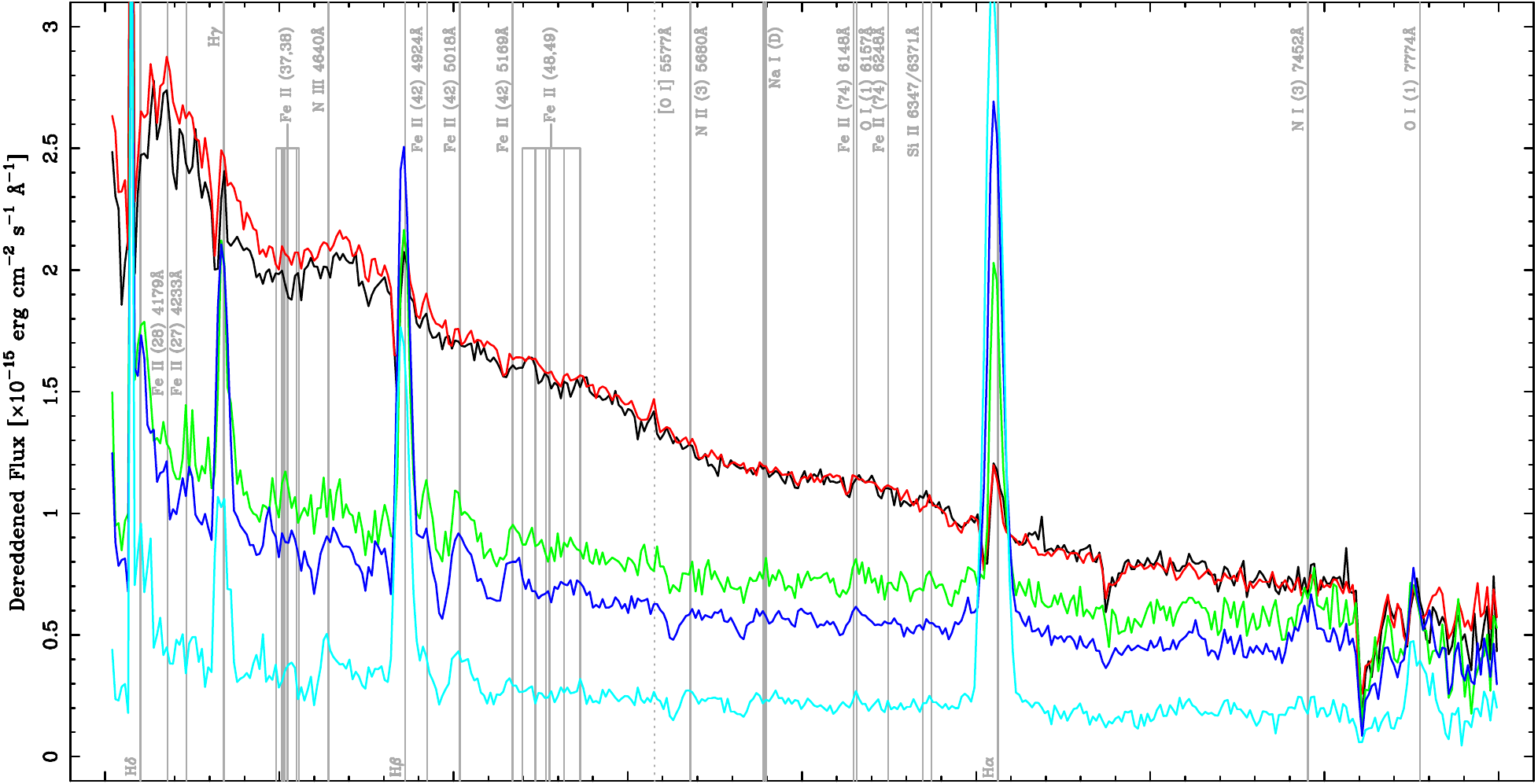}\\
\includegraphics[width=\textwidth]{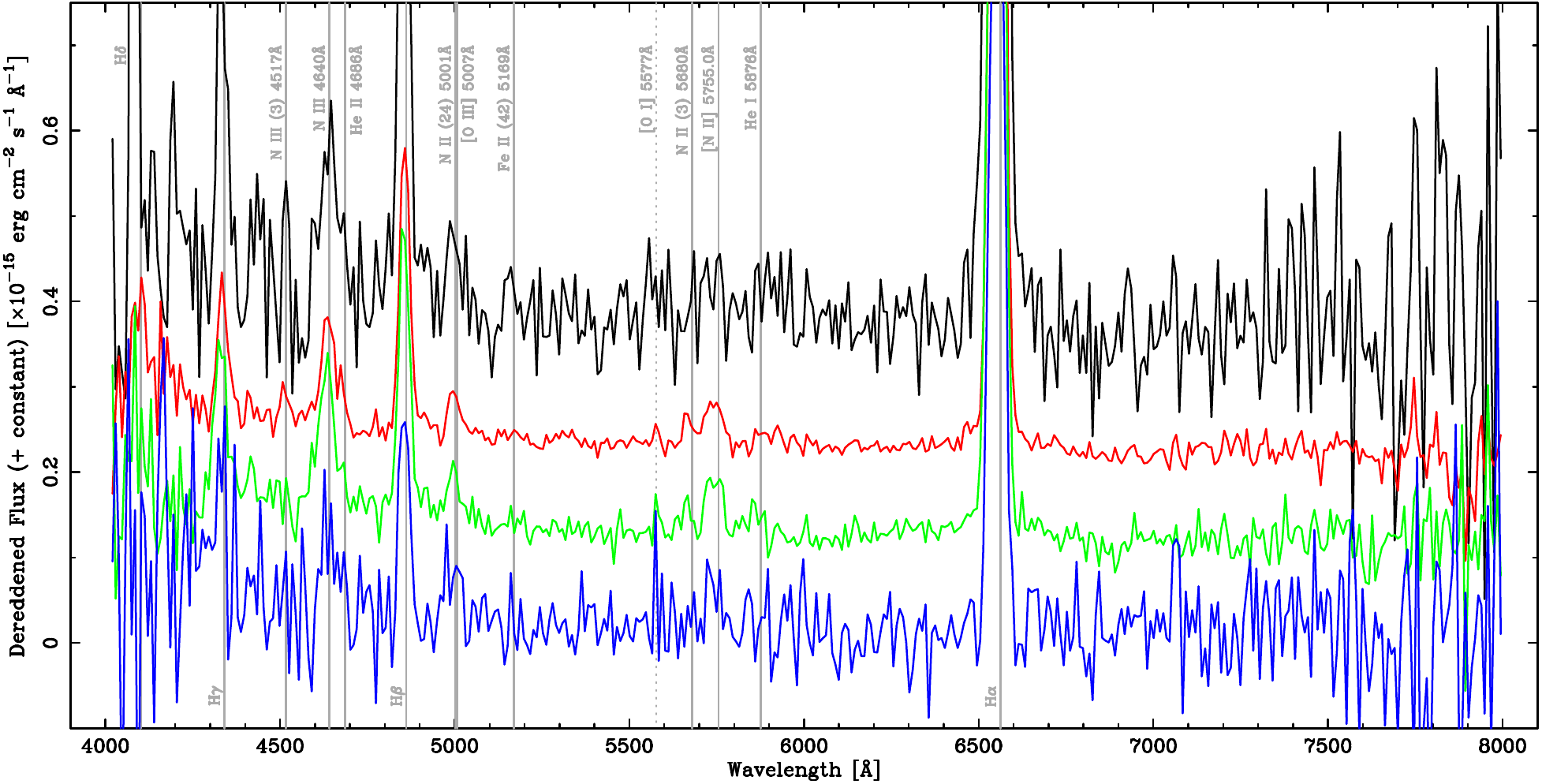}
\caption{The Liverpool Telescope SPRAT optical spectra of nova AT\,2016dah. {\bf Top:}\ Optically thick fireball and early decline `principle' Fe\,{\sc ii} spectra. Spectra from brightest to faintest:\ 2.150\,d post-eruption (black); 3.117\,d (red); 4.174\,d (green); and 7.141\,d (blue).  {\bf Bottom:}\ Transition from principle spectra to Orion phase. Brightest to faintest:\ 14.115\,d post-eruption (black); 22.133\,d (red); 24.149\,d (green); and 28.134\,d (blue). These four spectra have been offset in flux for clarity by integer multipliers of $0.1\times10^{-15}$\,erg\,cm$^{-2}$\,s$^{-1}$\,\AA$^{-1}$.\label{dah_spec}}
\end{figure*}

A series of thirteen spectra of AT\,2016dah were obtained by the SPRAT instrument on the LT. The first being captured at 2016 Jul 14.084\,UT, just 2.12\,d post-eruption. That spectrum, and a second obtained 0.05\,d later (see the black and red spectra -- the two most luminous -- in the top panel of Figure~\ref{dah_spec}), are both consistent with the early optically thick `fireball' phase of a nova eruption \citep{2016ATel.9248....1C}. These spectra were observed $\sim0.33$\,d before $r'$-band maximum light (see Figure~\ref{dah_lc}). This pair of spectra exhibit weak Balmer emission with the H$\alpha$--H$\delta$ line profiles all containing P\,Cygni absorption troughs to the blue. The H$\alpha$ absorption minimum is consistent with a velocity of $-1200\pm 200$\,km\,s$^{-1}$, with respect to the mean emission peak (see below). Whereas the terminal velocity of the H$\alpha$ P\,Cygni is at $2300\pm 400$\,km\,s$^{-1}$ blueward of the mean peak. These spectra display a blue continuum with relatively few other features. The continuum may be punctuated by blue-shifted absorptions from Fe\,{\sc ii} multiplets 42 and 74. At this stage emission from O\,{\sc i} (1) 7774\,\AA\ is already present. A fit to the continua of this pair of spectra shows that they are consistent with the form of a black body with effective temperature $\approx10,000$\,K.

\begin{figure*}
\includegraphics[width=\columnwidth]{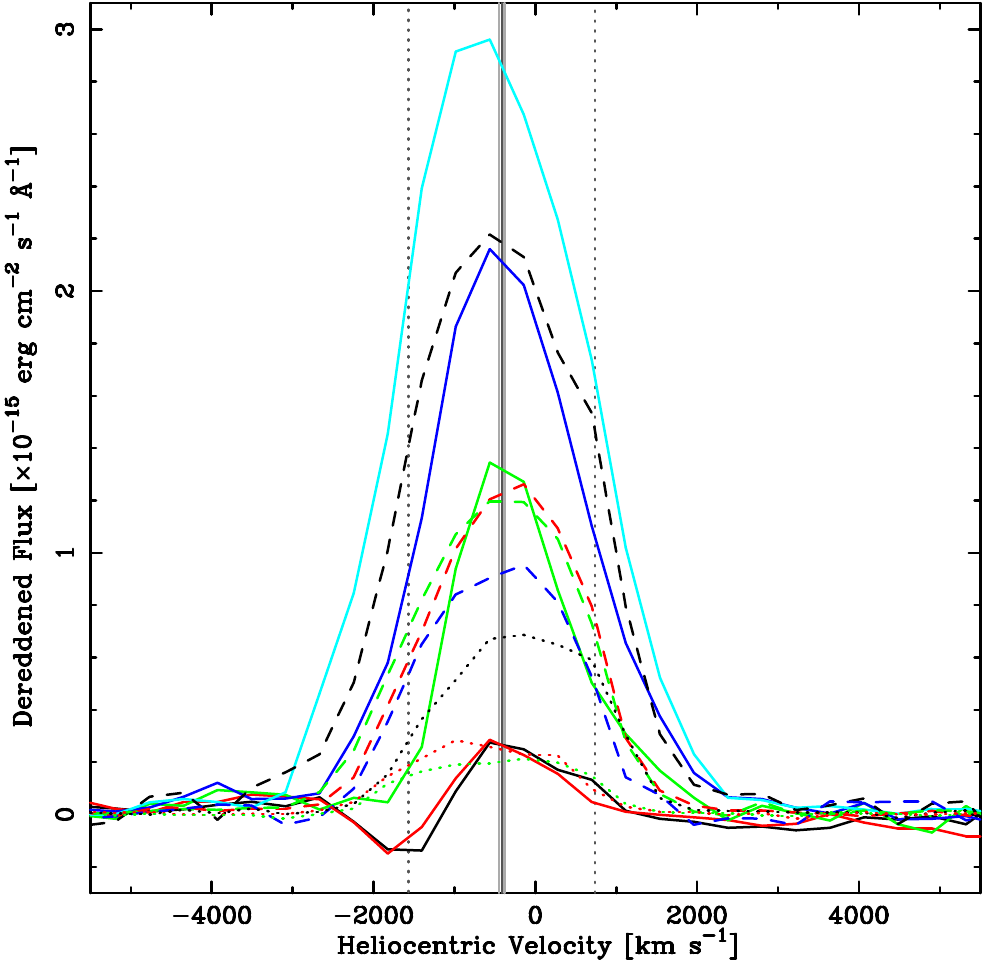}\hfill
\includegraphics[width=\columnwidth]{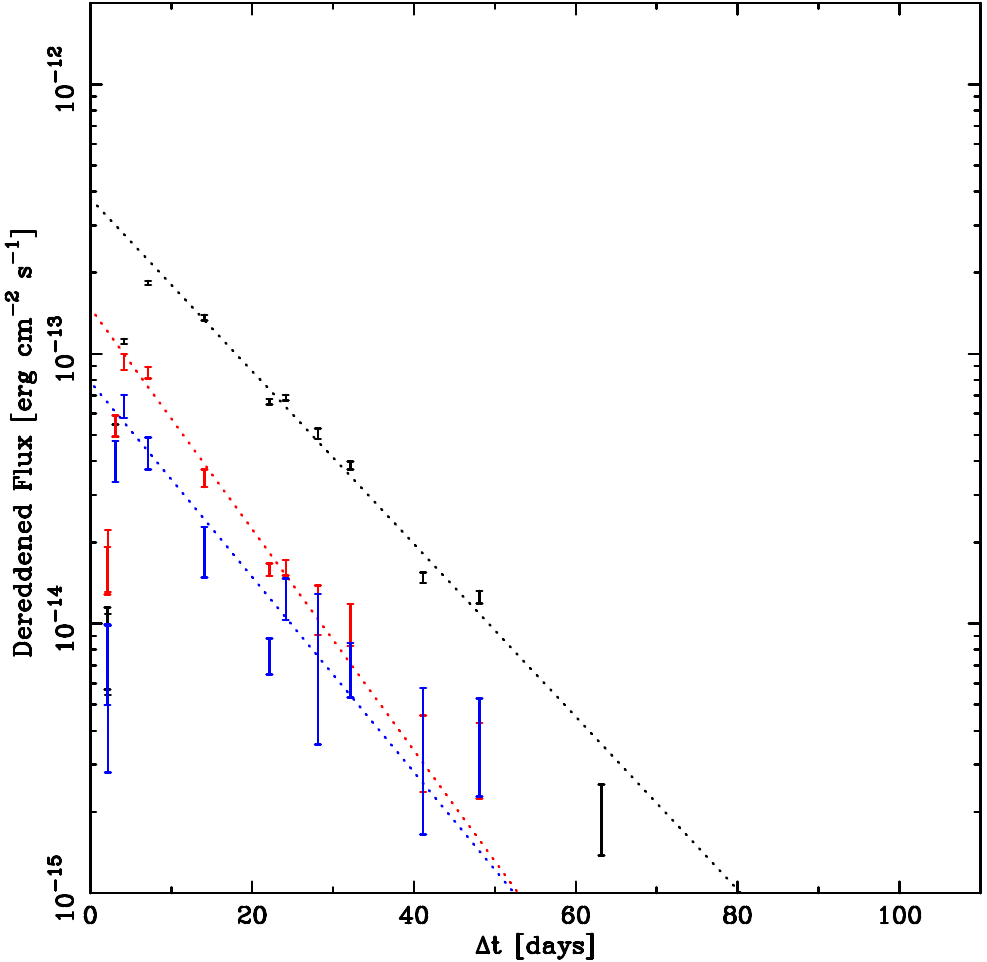}	\\
\includegraphics[width=\columnwidth]{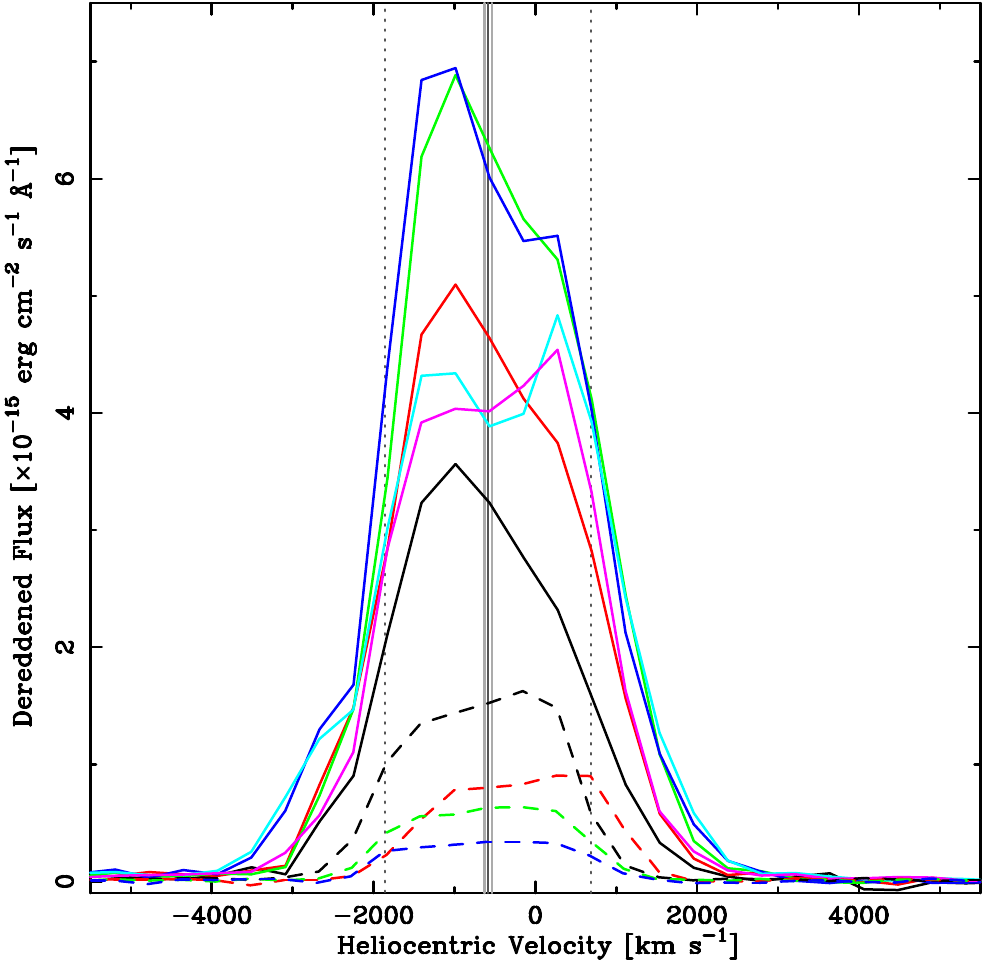}\hfill
\includegraphics[width=\columnwidth]{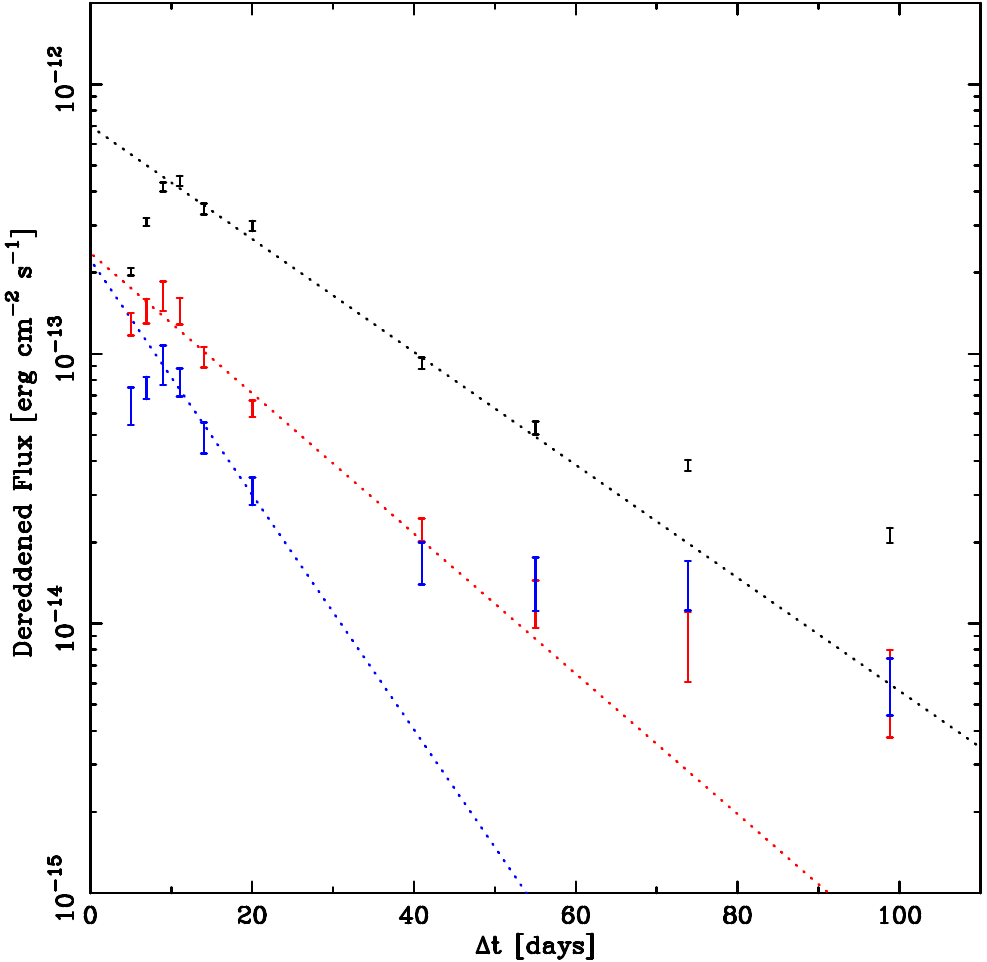}	
\caption{{\bf Left:}\ H$\alpha$ line profile evolution for AT\,2016dah (top) and AT\,2017fyp (bottom) following continuum subtraction and Heliocentric correction. The line colours are consistent with those used in Figures~\ref{dah_spec}, \ref{dah_spec2} and \ref{fyp_spec}. The solid and dashed lines relate to the spectra in the top and bottom panels of Figures~\ref{dah_spec}, \ref{fyp_spec}, respectively, the dotted lines (AT\,2016dah only) relate to those spectra shown in \ref{dah_spec2}. The vertical solid lines indicate the measured average line centre (and uncertainties), the vertical dotted lines indicate the extent of the measured average FWHM. {\bf Right:}\ The flux evolution of the H$\alpha$ (black), H$\beta$ (red), and H$\gamma$ emission lines for AT\,2016dah (top) and AT\,2017fyp (bottom). The dotted lines indicate linear fits to portions of those data (see text).\label{line_profiles}} 
\end{figure*}

A Gaussian profile combined with a linear function was used to simultaneously fit the line flux and local continuum for the Balmer emission lines (H$\alpha$--H$\gamma$; the low signal-to-noise and line blanketing, due to the low spectral resolution to the blue, rendered fits to H$\delta$ uninformative). The resultant line fluxes are reproduced in Table~\ref{flux}. The evolution of the Balmer emission line fluxes and that of the H$\alpha$ line profile are shown in the top-panel of Figure~\ref{line_profiles}. The Balmer emission peaks much later than the broad-band maximum light (which occurs at $\Delta t=2.48$\,days post-eruption). Following a slow increase in flux, the Balmer emission peaks around day 7 post-eruption. After peak, the Balmer emission follows a roughly linear decline (see Figure~\ref{line_profiles}). We measure the weighted average emission line centre of H$\alpha$ from all thirteen spectra to be at $-420\pm30$\,km\,s$^{-1}$ (after heliocentric correction) with respect to the rest-wavelength of H$\alpha$, which corresponds to a redshift of $z=\left(-1.4\pm0.1\right)\times10^{-3}$. The weighted average FWHM of H$\alpha$ across the thirteen spectra is $2300\pm70$\,km\,s$^{-1}$. Other than during the first two epochs, where the H$\alpha$ profile has a P\,Cygni form, there is not significant evolution of the FWHM through the spectral sequence. In general, the H$\alpha$ profile is Gaussian-like, which is fairly typical for Fe\,{\sc ii} novae.

\begin{table}
\caption{Selected emission line fluxes from the Liverpool Telescope SPRAT spectra of AT\,2016dah and AT\,2017fyp.\label{flux}}
\begin{center}
\begin{tabular}{llll}
\hline
& \multicolumn{3}{c}{Dereddened flux} \\
$\Delta t$\,[days] & \multicolumn{3}{c}{[$\times10^{-15}$\,erg\,cm$^{-2}$\,s$^{-1}$]}\\
\cline{2-4}
  & H$\alpha$ & H$\beta$ & H$\gamma$ \\
\hline
\multicolumn{4}{c}{AT\,2016dah} \\
\hline
\phantom{0}2.124$^\mathrm{a}$ & $\phantom{0}\phantom{0}8\pm3$ & $\phantom{0}16\pm3$& $\phantom{0}7\pm2$  \\
\phantom{0}2.175$^\mathrm{a}$ & $\phantom{0}\phantom{0}8\pm3$ & $\phantom{0}17\pm5$ & ($5\pm3$) \\
\phantom{0}3.117 & $\phantom{0}52\pm3$ & $\phantom{0}54\pm5$ & $40\pm7$ \\
\phantom{0}4.174 & $111\pm3$ & $\phantom{0}93\pm6$ & $64\pm6$ \\
 \phantom{0}7.141 & $183\pm3$ & $\phantom{0}85\pm4$ & $43\pm6$ \\
14.115 & $136\pm3$ & $\phantom{0}35\pm3$ & $18\pm4$ \\
 22.133 & $\phantom{0}66\pm1$ & $\phantom{0}16\pm1$ & $\phantom{0}8\pm1$ \\
24.149 & $\phantom{0}69\pm2$ & $\phantom{0}16\pm1$ & $12\pm2$ \\
28.134 & $\phantom{0}50\pm2$ & $\phantom{0}11\pm2$ & ($7\pm4$) \\
32.153 & $\phantom{0}39\pm1$ & $\phantom{0}10\pm2$ & $\phantom{0}7\pm2$ \\
41.133 & $\phantom{0}15\pm1$ & $\phantom{0}\phantom{0}3\pm1$ & ($3\pm2$) \\
48.090 & $\phantom{0}13\pm1$ & $\phantom{0}\phantom{0}3\pm1$ & ($3\pm1$) \\
63.167 & $\phantom{0}\phantom{0}2\pm1^\mathrm{b}$ & \phantom{0}\phantom{0}\ldots  & \ldots \\
\hline
\multicolumn{4}{c}{AT\,2017fyp}\\
\hline
\phantom{0}5.017 & $202\pm7$ & $130\pm10$ & $60\pm10$  \\
\phantom{0}6.938 & $310\pm10$ & $140\pm20$ & $75\pm7$\\
\phantom{0}9.001 & $420\pm20$ & $160\pm20$ & $90\pm20$\\
11.052 & $440\pm20$ &  $140\pm20$ & $78\pm9$\\
14.037 & $350\pm20$ & $100\pm9$ & $49\pm6$\\
20.025 & $300\pm10$ &  $\phantom{0}60\pm4$ & $31\pm4$ \\
40.965 & $\phantom{0}92\pm5$ &  $\phantom{0}22\pm2$ & $17\pm3$ \\
55.046 & $\phantom{0}53\pm3$ & $\phantom{0}12\pm2$ & $14\pm3$ \\
73.859 & $\phantom{0}39\pm2$ & $\phantom{0}\phantom{0}8\pm2$ & $14\pm3$ \\
98.832 & $\phantom{0}21\pm1$ & $\phantom{0}\phantom{0}5\pm2$ & $\phantom{0}6\pm1$ \\
\hline
\end{tabular}
\end{center}
\begin{tablenotes}
\small
\item All emission line fluxes and uncertainties were determined through fitting a Gaussian profile to the line in velocity space. Values in parenthesis are lines for which the flux uncertainties lie between 2 and $3\sigma$.
\item $^\mathrm{a}$In these spectra the Balmer lines have P\,Cygni profiles, here we report the flux of the emission component.
\item $^\mathrm{b}$Formally, the flux was constrained marginally beyond $3\sigma$.
\end{tablenotes}
\end{table}

The following three spectra were taken, 3.12\,d, 4.17\,d, and 7.14\,d post-eruption (see green, blue, and cyan, respectively, spectra in the top panel of Figure~\ref{dah_spec}) and were all collected between maximum light and $t_2$, with the last of this sub-set taken just over 1 magnitude below peak. Here there is a clear transition from the fireball phase to the `principle' spectrum. The continuum emission weakens across the sub-set, it is no longer black-body like, and is shallower than that expected from a Rayleigh-Jeans tail. Throughout this stage the continuum slope is not consistent with that expected from free-free emission \citep[see][]{1975MNRAS.170...41W}. The Balmer series has transitioned from their initial P\,Cygni profiles to strong emission lines that contribute substantially to the total optical emission. The Fe\,{\sc ii} (42) triplet is the next strongest feature, with emission also detected from Fe\,{\sc ii} (26, 27, 37, 38, 48, 49, 74). Given the lack of Fe\,{\sc ii} (74) 6248\,\AA, the Fe\,{\sc ii} (74) 6148\,\AA\ line is probably blended with O\,{\sc i} 6157\,\AA. Emission from O\,{\sc i} (1) 7774\,\AA\ remains present and is joined by weak Si\,{\sc ii} 6347/6371\,\AA\ and Na\,{\sc i} (D) lines.  Emission from N\,{\sc i} (3) at $\approx7452$\,\AA\ also becomes visible at this time. These spectra are consistent with the nova being a member of the Fe\,{\sc ii} taxonomic spectral class \citep{1992AJ....104..725W,2012AJ....144...98W}. The latter spectra of this sub-set show evidence for the appearance of emission from N\,{\sc iii} 4640\,\AA\ (from day 4.17) and N\,{\sc ii} (3) 5680\,\AA\ (from day 7.14). The apparent presence of [O\,{\sc i}] 5577\,\AA\ is most likely the residual from sky-subtraction.

The next set of four spectra were collected between two-weeks and four-weeks post-eruption (see the bottom panel of Figure~\ref{dah_spec}) and span the epoch of $t_2$ to approximately $t_3$. Here the systemic fading and flattening of the continuum has continued and the Balmer emission has weakened. The Fe\,{\sc ii} (42) emission remains at day 14.12, but no Fe\,{\sc ii} emission is evident from day 22 onward. Could the `drop' seen in the $B$ and $V$-band light curves around day 14 (see Section~\ref{dah_phot}) be solely due to the weakening of the Fe\,{\sc ii} emission, as the Balmer decline is smooth during this phase (see Figure~\ref{line_profiles})? Emission lines of N\,{\sc ii} 5001, 5680\,\AA\ and N\,{\sc iii} 4517, 4640\,\AA\ are visible throughout this set \citep[also see][]{2016ATel.9329....1C}. The N\,{\sc ii} 5001\,\AA\ is seen to visibly strengthen compared to the neighbouring H$\beta$ line. There is evidence for the appearance of [N\,{\sc ii}] 5755\,\AA\ from day 14, along with He\,{\sc i} 5876\,\AA. The diminishing signal-to-noise in these spectra obscures any O\,{\sc i} line that may still be present. There is tentative evidence for the He\,{\sc ii} 4686\,\AA\ line appearing from day 14. The movement of the flux ratio of H$\beta$:N\,{\sc ii} 5001\,\AA\ toward the N line (around half the H$\beta$ flux by day 28) could also suggest that the [O\,{\sc iii}] 5007\,\AA\ nebular line may be be present and strengthening.

\begin{figure*}
\includegraphics[width=\textwidth]{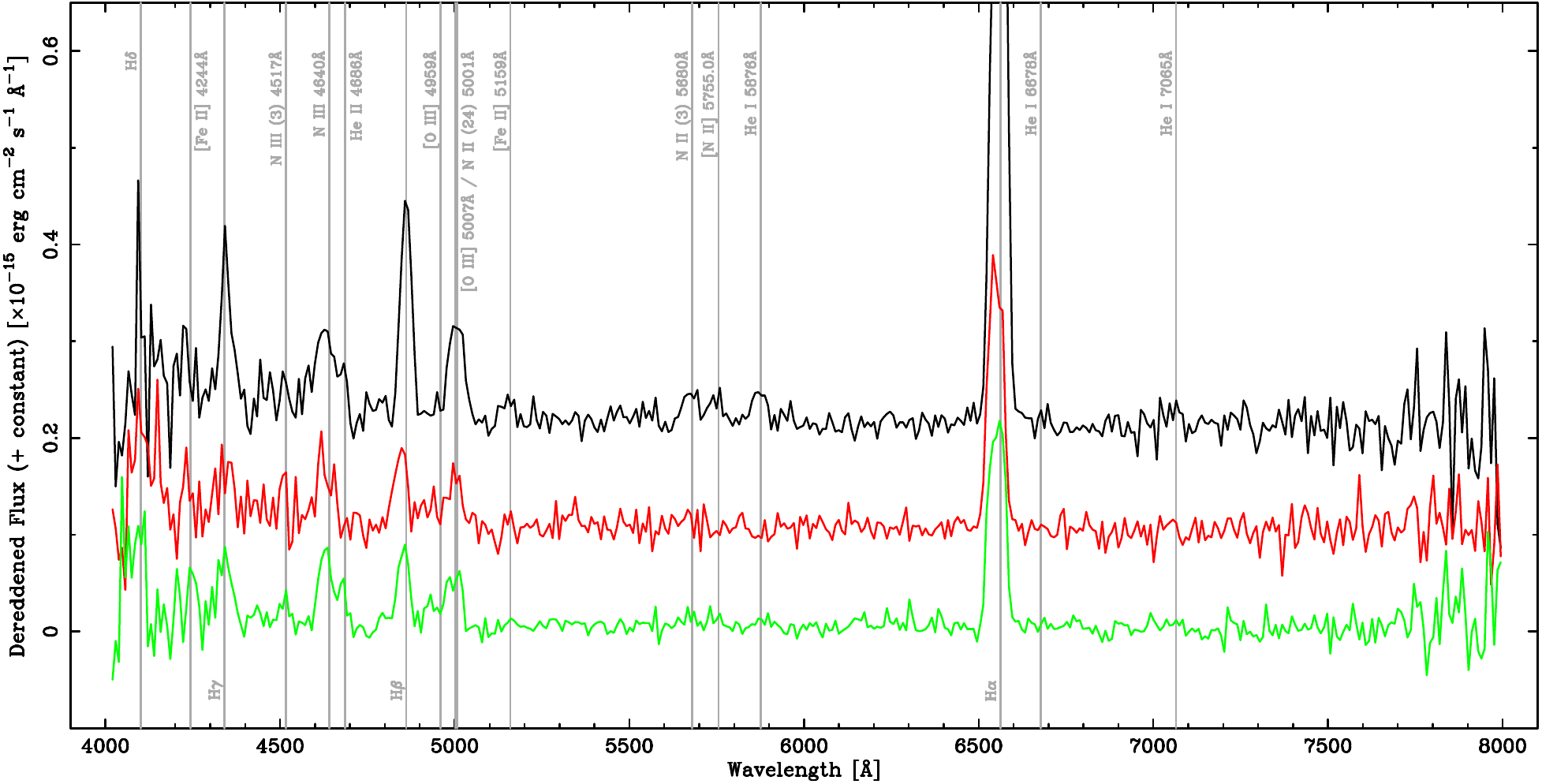}
\caption{The Liverpool Telescope SPRAT optical spectra of nova AT\,2016dah. These spectra show the nebular phase of the eruption. Brightest to faintest:\ 32.153\,d post-eruption (black); 41.133\,d (red); and 48.090\,d (green). These three spectra have been offset in flux for clarity by integer multipliers of $0.1\times10^{-15}$\,erg\,cm$^{-2}$\,s$^{-1}$\,\AA$^{-1}$. The spectrum from 2016 Sep 13 ($\Delta t=63.167$\,d) is not shown due to its low signal-to-noise.\label{dah_spec2}}
\end{figure*}

The final sub-set of four spectra (see Figure~\ref{dah_spec2}) were taken between day 32 and 63 post-eruption, these range from just post-$t_3$ to around five magnitudes below peak. The continuum continues to fade, remaining just detected in the final spectrum, the Balmer flux continues to fade. Other prominent lines include the N\,{\sc iii} 4640\,\AA\ feature (the Bowen-blend), and a line at around 5000\,\AA\ that rivals the flux of H$\beta$ and is most likely [O\,{\sc iii}] 5007\,\AA. Other lines present include He\,{\sc i} 5876, 6678, 7065\,\AA, He\,{\sc ii} 4686\,\AA, N\,{\sc ii} 5680\,\AA\, [N\,{\sc ii}] 5755\,\AA, N\,{\sc iii} (3) 4517\,\AA, and possibly [Fe\,{\sc ii}] 4244, 5159\,\AA, and [O\,{\sc iii}] 4959\,\AA. As such, AT\,2016dah becomes one of a small handful of extragalactic novae to have been detected in their nebular phase.

\subsubsection{AT\,2017fyp}

\begin{figure*}
\includegraphics[width=\textwidth]{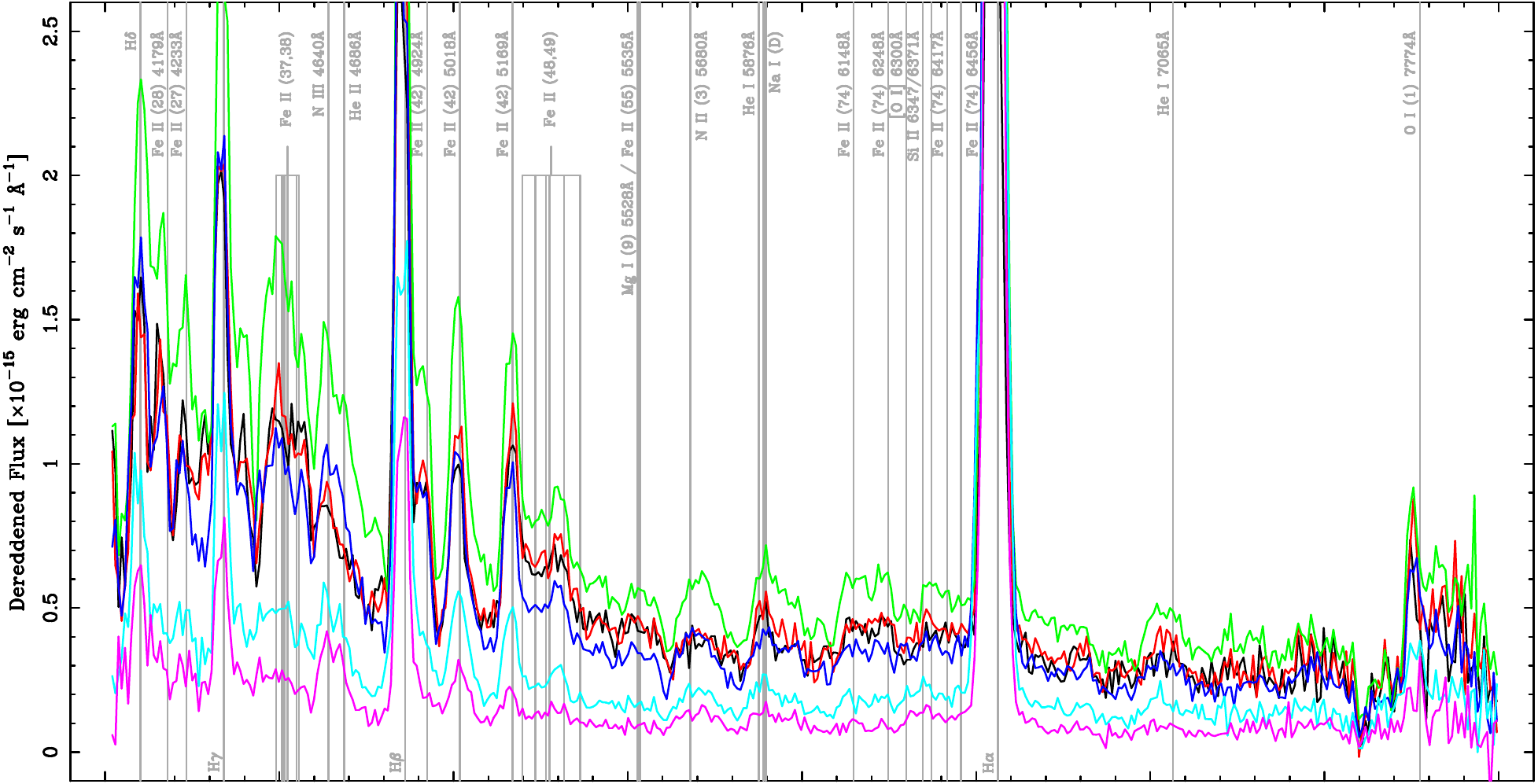}\\
\includegraphics[width=\textwidth]{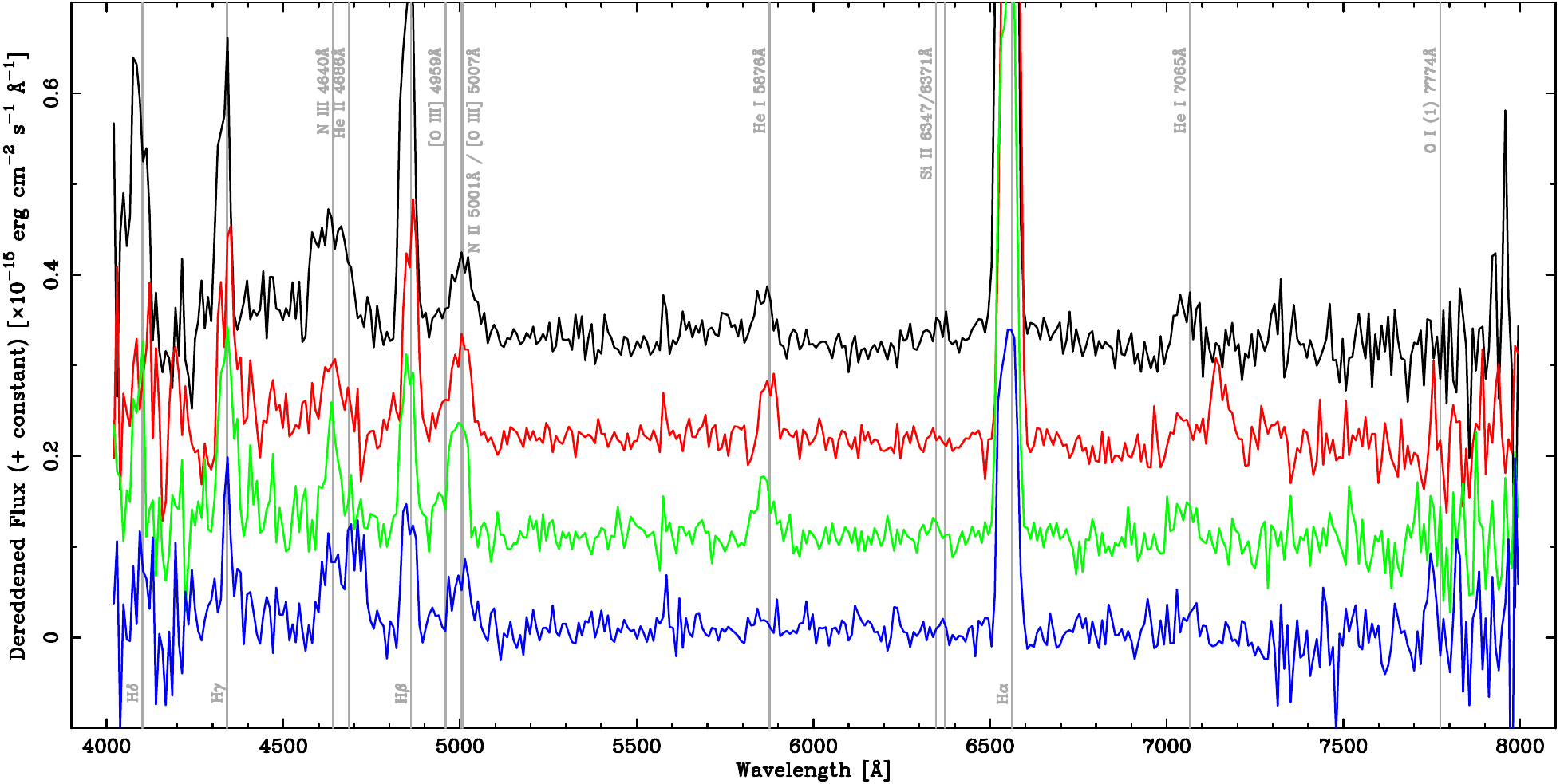}
\caption{The Liverpool Telescope SPRAT optical spectra of nova AT\,2017fyp. {\bf Top:}\ The principle Fe\,{\sc ii}+He/N (hybrid) spectra during the early decline spectra. Spectral sequence:\ 5.017\,d post-eruption (black); 6.938\,d (red); 9.001\,d (green); 11.052\,d (blue); 14.037\,d (cyan); and 20.025\,d (magenta).  {\bf Bottom:}\ Transition from Orion to nebular spectra. Brightest to faintest:\ 40.965\,d post-eruption (black); 55.046\,d (red); 73.859\,d (green); and 98.832\,d (blue). These four spectra have been offset in flux for clarity by integer multipliers of $0.1\times10^{-15}$\,erg\,cm$^{-2}$\,s$^{-1}$\,\AA$^{-1}$.\label{fyp_spec}}
\end{figure*}

The spectral coverage of AT\,2017fyp began later than that of AT\,2016dah. The first spectrum of this eruption was obtained 5\,days post eruption and began a series of eleven spectra all collected by SPRAT on the LT. As discussed in Section~\ref{fyp_phot}, the light curve of AT\,2017fyp either indicates a slow rise to peak, or alternatively an approximately flat topped profile. In either event the evolution during the first three spectra, taken on days 5, 7, and 9 (black, red, and green spectra in the top panel of Figure~\ref{fyp_spec}), is consistent with the slight increase in photometric flux (particularly in the blue) seen during this stage. These three and the remaining three in this sub-set, which were taken 11, 14, and 20 days post-eruption (the blue, cyan, and magenta spectra, respectively), were all collected between maximum light and approximately $t_2$. Following the initial rise in continuum luminosity there follows a systemic decline in flux. All these spectra are dominated by Balmer emission, and there is no evidence for absorption components in the earlier spectra. This would seem to suggest that the light curve is indeed flat topped, rather than this stage being a slow rise to maximum (with the continuum still optically thick). The width of the H$\alpha$ line in the first epoch is $2440\pm60$\,km\,s$^{-1}$. Emission lines from Fe\,{\sc ii} (42) are particularly prominent; those from multiplets 26, 27, 37, 38, 48, 49, and 74 are also present. A line at $\sim5530$\,\AA\ may be due to Fe\,{\sc ii} (55) or could be Mg\,{\sc i} (9) 5528\,\AA. The O\,{\sc i} (1) 7774\,\AA\ line is present as is the Si\,{\sc ii} doublet. However, throughout this whole sub-set the following lines are also clearly present:\ He\,{\sc i} 5876, 7065\,\AA, N\,{\sc ii} (3) 5680\,\AA, N\,{\sc iii} 4640\,\AA, and possibly He\,{\sc ii} 4686\,\AA\ \citep[although, at this early stage, this line may actually be O\,{\sc ii} 4650\,\AA; see, for e.g.,][]{2018A&A...611A...3H}. As such, these spectra best fit the hybrid taxonomic class \citep[also see][]{2017ATel10628....1H}; although a good number of hybrids display a transition between spectral types, rather than displaying both so prominently for an extended period. Over this period the slope of the continuum (albeit hard to define unambiguously) appears roughly consistent with that expected for optically thin free-free emission, as early as day 5 post-eruption.

Gaussian fits were again performed for the H$\alpha$--H$\gamma$ emission lines. The evolution of the Balmer line fluxes and the H$\alpha$ line profile is shown in the bottom two panels of Figure~\ref{line_profiles}. As with AT\,2016dah, the Balmer emission peaks after maximum light (which we assume occurred between $4\lesssim\Delta t\lesssim9$\,d post-eruption). Again, following a slow rise, the Balmer emission peaks at $\approx11$\,d post-eruption. After this peak, the Balmer line emission declines with an approximately linear form until $\sim50$ days post-eruption where the decay rate seems to slow. The weighted average of the Heliocentric corrected line centre of H$\alpha$ from all ten spectra is $-580\pm50$\,km\,s$^{-1}$, which corresponds to a redshift of $z=\left(-1.9\pm0.2\right)\times10^{-3}$. The H$\alpha$ line profile of AT\,2017fyp has the `boxy' form that is typical of He/N novae.  There is no significant evolution in the line width across all ten spectra, the (weighted) average H$\alpha$ width is $2550\pm60$\,km\,s$^{-1}$.

The final sub-set of four spectra were collected between day 41 and 99 post-eruption (the black, red\footnote{The apparent broad line, redward of He\,{\sc i} in the $\Delta t=55$\,d spectrum, is actually the ghost image of a bright and saturated star that has persisted from a previous spectrum acquisition observation performed by SPRAT.}, green, and blue spectra, respectively, in the bottom panel of Figure~\ref{fyp_spec}), and range from $\sim t_3$ throughout the late-decline. In all four, the continuum emission appears essentially consistent, but is only marginally detected. These spectra remain dominated by the waning Balmer emission, by day 41, any Fe\,{\sc ii} emission is undetected. The most prominent lines are the Bowen-blend complex and [O\,{\sc iii}] 5007\,\AA\ (again supported by the waning strength of H$\beta$ relative to this line and the 4959\,\AA\ line). He\,{\sc ii} 4686\,\AA\ also seems to be present. He\,{\sc i} 5876, 7065\,\AA\ remain detected, as do O\,{\sc i} (1) 7774\,\AA\ and the Si\,{\sc ii} doublet (at day 41). This indicates that, as was the case for AT\,2016dah, we have observed AT\,2017fyp in its nebular phase.

\section{Spatial Distribution}\label{sec:spatial}

The Andromeda Stream, or the Southern Andromeda Stream, or the M\,31 Giant/Great Stellar Steam was discovered by \citet[also see \citealt{2002AJ....124.1452F}]{2001Natur.412...49I} who employed a wide-field survey of the halo of M\,31 using the Isaac Newton Telescope. The GSS appears almost linear on the sky and roughly follows a line connecting M\,32 and M\,110 \citep[aka NGC\,205;][]{2001Natur.412...49I}, extending to $\sim5^\circ\!\!.5$ to the south of M\,31 and $\sim3^\circ\!\!.5$ to the east \citep[see their Figure~23, which also indicates the main structures around M\,31]{2007ApJ...671.1591I}. The GSS ranges from $\sim100$\,kpc behind M\,31 at its southern-most extreme to around 30\,kpc in front of M\,31 at its northern reach \citep{2003MNRAS.343.1335M}. Modelling by \citet{2006AJ....131.1436F} indicated that the mass of the stream's progenitor (a satellite of M\,31) $>10^8$\,M$_\odot$ -- a massive dwarf galaxy. \citet{2004MNRAS.351..117I} and \citet{2006AJ....131.1436F} used velocity arguments to exclude an M\,32 or M\,110 passage of M\,31 as the source of the stream. The latter proposed Andromeda\,VIII \citep{2003ApJ...596L.183M} as the potential progenitor, whereas the former suggested that And\,VIII may simply be part of the stream, a suggestion backed up by \citet{2006MNRAS.369..120M}. More recently, \citet{2013MNRAS.434.2779F} proposed that the progenitor was a `previously unknown' M\,31 satellite with stellar mass $\left(3.7\pm0.7\right)\times10^9$\,M$_\odot$, of order the mass of the Magellanic Clouds, and that the encounter with M\,31 occurred 760$\pm50$\,Myr ago.

\begin{figure}
\includegraphics[width=\columnwidth]{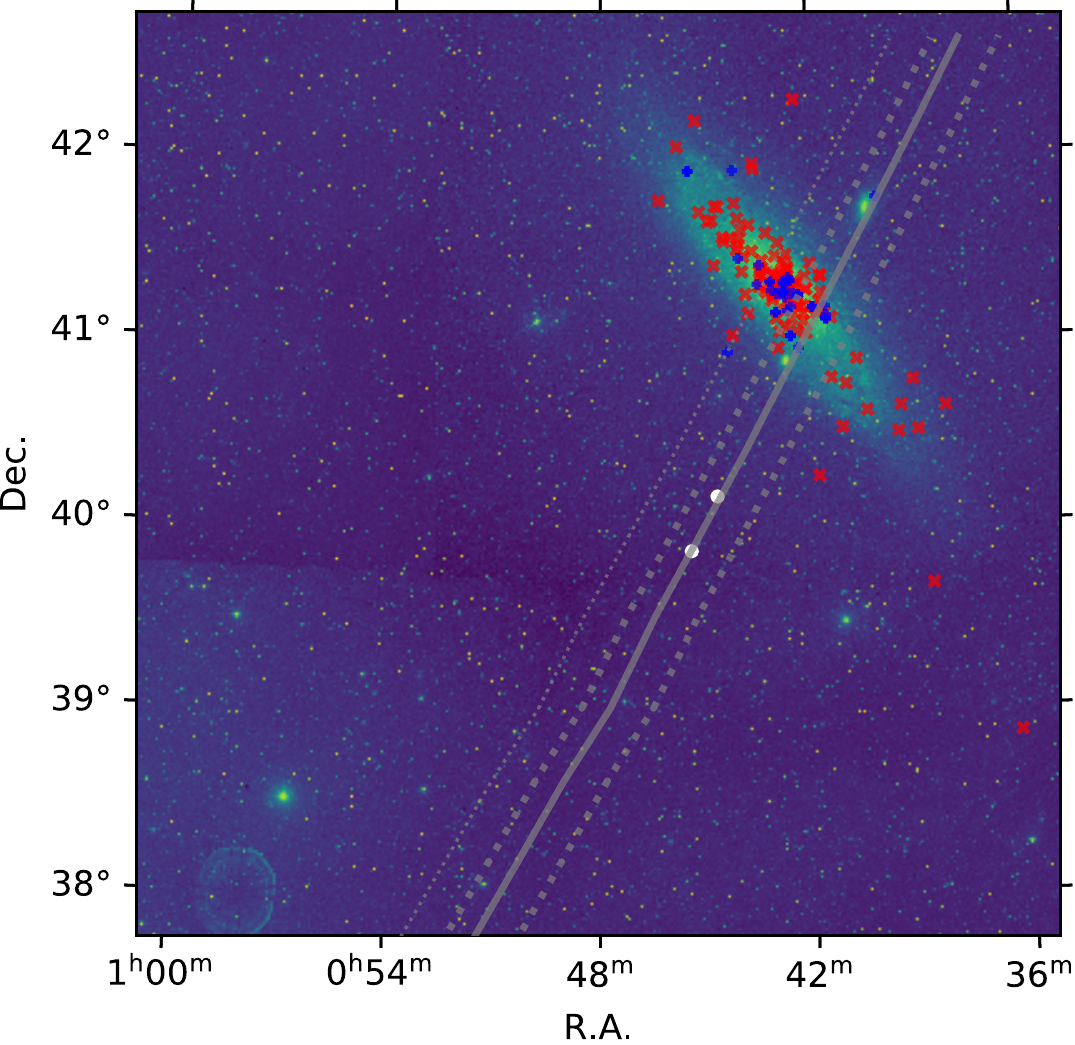}
\caption{As Figure~\ref{M31_field}, but showing a much wider field, with M\,31 offset to the north-west. The solid grey diagonal line indicated the approximate location of the peak stellar density of the M\,31 Giant Stellar Stream, the dashed grey lines delimits the bulk of the stream's stellar content, the dotted grey line indicates the lower density eastern confines of the stream. By inspection, AT\,2016dah and AT\,2017fyp clearly lie along the stream's central peak density region.\label{stream_img}}
\end{figure}

In Figure~\ref{stream_img} we partially recreate those data shown initially in Figure~\ref{M31_field}. However, the field of view has been extended from $3^\circ\times3^\circ$ to $5^\circ\times5^\circ$, with M\,31 located in the north-west (top--right). Again, all M\,31 novae with known spectral types from \citet{2011ApJ...734...12S} and C.\ Ransome et al.\ (in preparation) are shown, as are the locations of AT\,2016dah and AT\,2017fyp, to the far south. We have over-plotted the locus of the M\,31 GSS based upon the descriptions within the literature, particularly \citet{2001Natur.412...49I} and \citet{2003MNRAS.343.1335M}. Here the solid grey line indicates the approximate line of peak stellar density along the stream, the dashed lines indicate the western edge and the region of equivalent stellar density to the east, the dotted line indicates the eastern extreme (also see Figure~\ref{stream_modelapprox}). From inspection alone, the alignment of AT\,2016dah and AT\,2017fyp with the central line of the M\,31 GSS is remarkable. 

The location of both AT\,2016dah and AT\,2017fyp along the line of the GSS (see Figure~\ref{stream_img}) strongly implies that the novae are physically located within the GSS, but with the spatial extent of surveys of M\,31 and the substantial number of novae detected, it is important to consider the possibility that this alignment is a chance coincidence, and the novae are unconnected with the GSS. 

\subsection{Modelling the spatial distribution of a sample of novae}\label{sec:spatial_model}

We therefore require a model for the underlying distribution of nova in our sample, assuming that the GSS is not a source of novae. Armed with such a model, we can create Monte Carlo (MC) simulations of the nova distribution and explore that fraction that have a similar (or greater) association to the GSS. We adopt the approach of \citet{2006MNRAS.369..257D} and \citet[also see \citealt{1987ApJ...318..520C}, \citealt{2001ApJ...563..749S}, and \citealt{2019arXiv190910497D}]{2016ApJ...817..143W} who define:

\begin{equation}
\Psi_i=\frac{\theta\mathscr{L}_i^d+\mathscr{L}_i^b}{\theta\sum_i\mathscr{L}_i^d+\sum_i\mathscr{L}_i^b},
\label{nova_mod}
\end{equation}

\noindent where, over a grid of positions $i$, $\Psi$ is the probability of a nova erupting at a given location, $\theta$ is the ratio of the disk and bulge nova eruption rates per unit ($r'$-band) flux, and $\mathscr{L}_i^b$ and $\mathscr{L}_i^d$ the disk and bulge contributions to the ($r'$-band) flux at that location, respectively. Using this model \citet{2006MNRAS.369..257D} studied a sample of novae \citep[see][]{2004MNRAS.353..571D} close to the core of M\,31 and found $\theta=0.18$. \citet{2016ApJ...817..143W} adopted the same value of $\theta$ but, with a larger, more heterogenous sample, they found that a correction needed to be applied to compensate for variable completeness of the sample.

To estimate the flux component $\mathscr{L}_i^b$ we model the bulge with elliptical isophotes with an axial ratio $b/a = 0.6$ \citep{1987ApJ...318..520C} and a standard $r^{1/4}$ law \citep{1953MNRAS.113.134}. For the disk light $\mathscr{L}_i^d$, an exponential is used \citep{1970ApJ.160.811}. Both of these models are defined along the semi-major axis of the component ($a_b$ and $a_d$ for bulge and disk, respectively) yielding:

\begin{equation}
\log \left  [ {\mathscr{L}_b\left(a_b\right)/\mathscr{L}_b^0} \right ] = -3.33 \left [ { \left ({a_b/a_b^0}\right )^{1/4} - 1 }\right ],
\end{equation}

\noindent where $\mathscr{L}_b$ is the flux at some distance $a_b^0$ and is used for normalisation, and

\begin{equation}
\mathscr{L}_d(a_d) = \mathscr{L}_d^0 \, \mathrm{e}^{-a_d/a_d^0},
\end{equation}

\noindent where $\mathscr{L}_d^0$ and $a_d^0$ are similarly normalisation factors.

We adopt a similar approach, but, clearly, require a different correction factor for our sample, which are the spectroscopically confirmed novae in the samples from \citet{2011ApJ...734...12S} and C.\ Ransome et al.\ (in preparation), as shown in Figure~\ref{M31_field}. It is important to note that the purpose of this model is to produce suitable MC simulations that mirror the underlying constraints of the observed sample. There is a degeneracy between the specific value for $\theta$ used and the details of the completeness correction, so no conclusions can be drawn from the exact values of either, but as long as the observed distribution is matched, the MC simulations can be used to test the hypothesis that the GSS is not the source of any novae. 

For each MC iteration, the model is used to create a probability of a nova occurring at any observed position in a fine grid, with a resolution of $4^{\prime\prime}$ over a large ($10^\circ \times 10^\circ$) field, and a random number generator used to seed a single nova. The simulated nova is associated at random to either the disk or bulge (weighted appropriately by the model flux values at that point) and a de-projected radius $r$ is calculated (i.e.\ the equivalent semi-major axis distance). The completion factor at that radius is then used to determine whether that particular nova is ``observed'' and placed into the mock catalogue for that iteration. This is repeated until the number of novae in the MC iteration matches the observed sample (276). This process is in turn repeated for a large number of MC iterations (at least 1000) to produce at least 1000 separate simulated nova ``catalogues''.

For each MC iteration, a set of $r$ values for the observed nova are generated. Since the value of $r$ for a particular nova depends upon whether it is in the bulge or disk, and that is not known for the novae, the model is used to associate a probability of being bulge or disk with, for each MC iteration, every observed nova assigned at random in line with that probability and hence an appropriate $r$ calculated. Since the MC simulations only require us to produce a distribution with the correct radial characteristics, we do not need to estimate or model the full on-sky completeness function, but just correct for the incompleteness in $r$. Therefore, the completion correction is designed to minimise the difference between the cumulative distribution in $r$ of real and simulated nova. It was generated by taking the ratio of the cumulative distributions of uncorrected and observed nova in a series of bins in $\log(r)$ and fitting a set of steps to the ratio.

\begin{figure}
\includegraphics[width=\columnwidth]{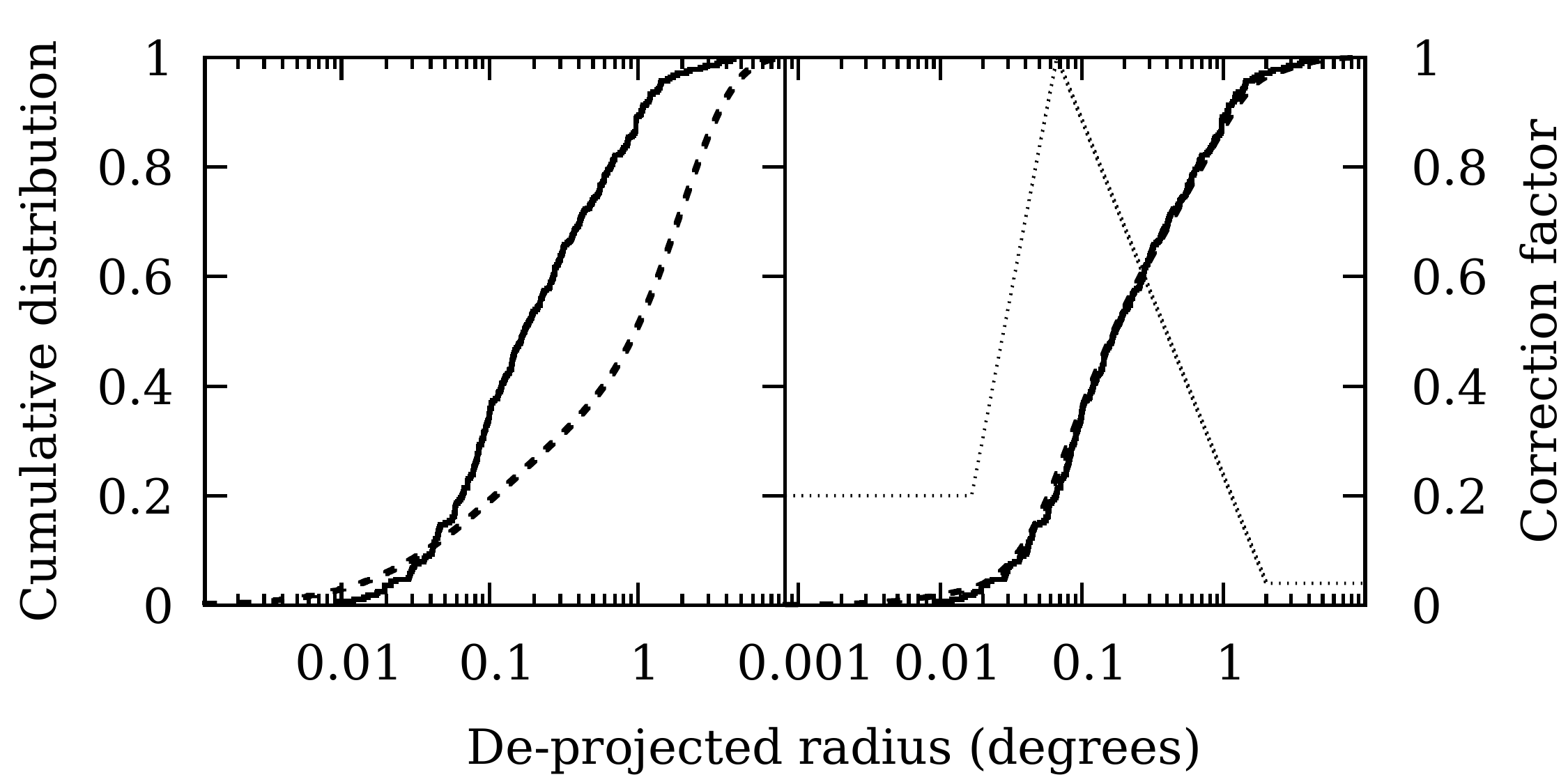}
\caption{Comparison of the spatial distribution model to the observed nova sample. {\bf Left:}\ the cumulative distribution in $r$ of the observed novae (solid line) is compared to the equivalent distribution from a combination of 1000 MC iterations with a model value of $\theta=0.18$ (dashed line), without any completion correction. {\bf Right:}\ the same observed distribution is compared to the result from MC simulations where the four-step correction function shown (thin, dotted line) is applied. A \citet{kolmogorov33}--\citet{smirnov1948} test applied to the two sets of distributions gives a 0\% chance that the uncorrected simulations are drawn from the observed distribution of novae, but a 67\% chance that the corrected simulations are drawn from the observed distribution.
\label{spatial_radcorr}}
\end{figure}

This can be seen in Figure~\ref{spatial_radcorr} where the cumulative distribution in $r$ of the observed nova sample and combined MC iterations are compared. The correction factor comes in four steps: close to the core, completion is low; completion then increases out into the disk and is normalised to $1$ at its maximum before dropping down to a low level. This matches what might be expected. In the inner regions, the high background makes detection and spectroscopic follow-up of novae difficult, so as the flux from the galaxy drops, the completeness increases. However, further out coverage by surveys is less complete (particularly until recently, also see Section~\ref{sec:stream}) and so completion drops again. However, it is important to remember that, although the form of the completion correction seems sensible, it is only used ensure that the model reproduces the observed distribution:\ a different value of $\theta$ would result in a different correction function, but the overall results would be largely unchanged.

\subsection{Modelling the Stream}\label{sec:spatial_strream}

\citet{2003MNRAS.343.1335M} provide a detailed description of the stellar light from the GSS, which can be used to produce a simplified model that we can apply to our observed nova sample and MC simulations. Our model follows the centre line of the GSS as determined by \citet[also see Figure~\ref{stream_img}]{2003MNRAS.343.1335M}, but approximates the light with a two-dimensional representation along the GSS.

\begin{figure}
\includegraphics[width=\columnwidth]{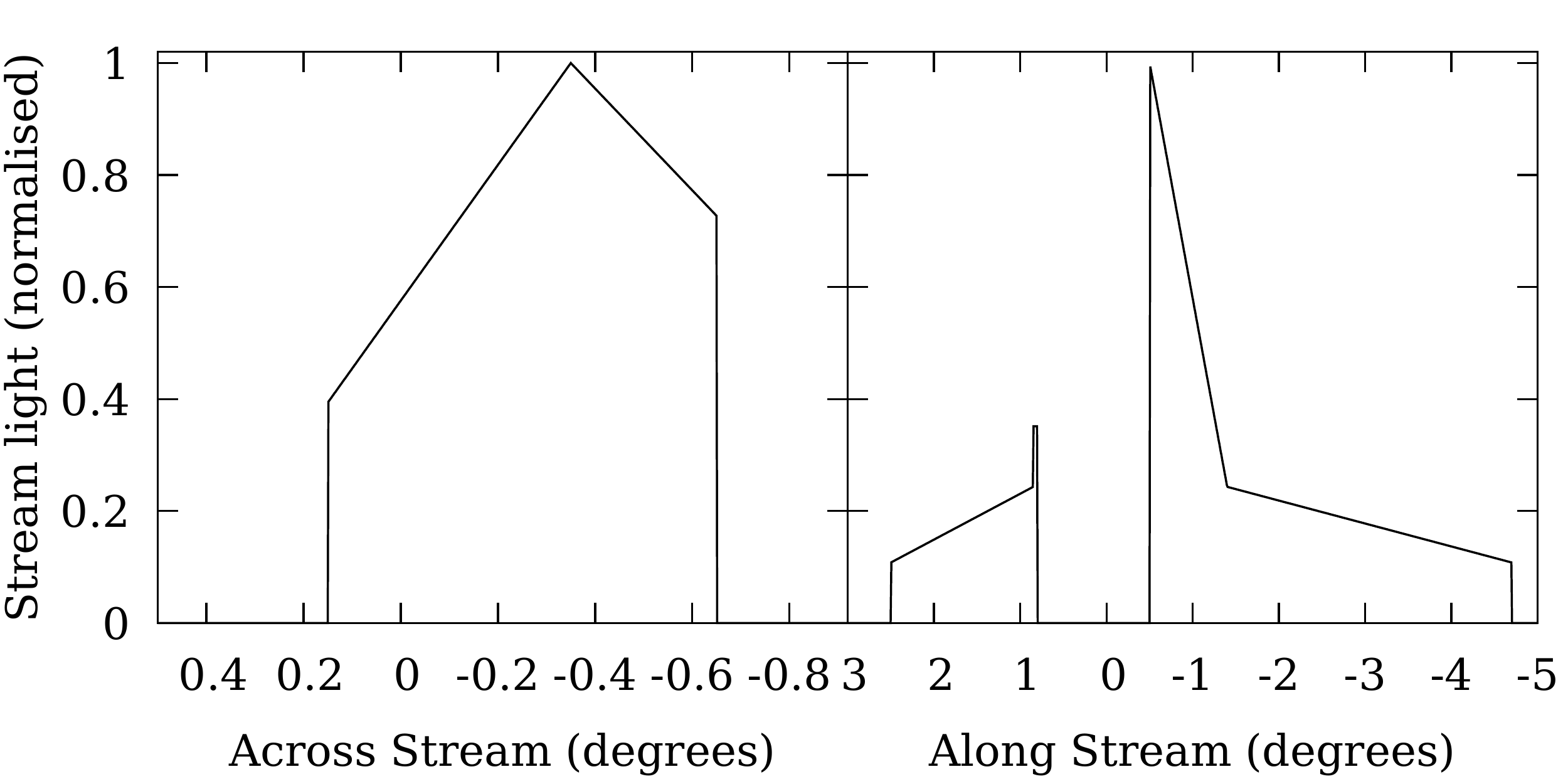}
\caption{Our approximation to the GSS stellar light originally presented by \citet{2003MNRAS.343.1335M}. On the left is simplified functional form we use to determine the average light along the GSS and on the right the equivalent across the GSS. The model light at any point is the product of the two functions at that position.
\label{stream_modelapprox}}
\end{figure}

\begin{figure}
\includegraphics[width=\columnwidth]{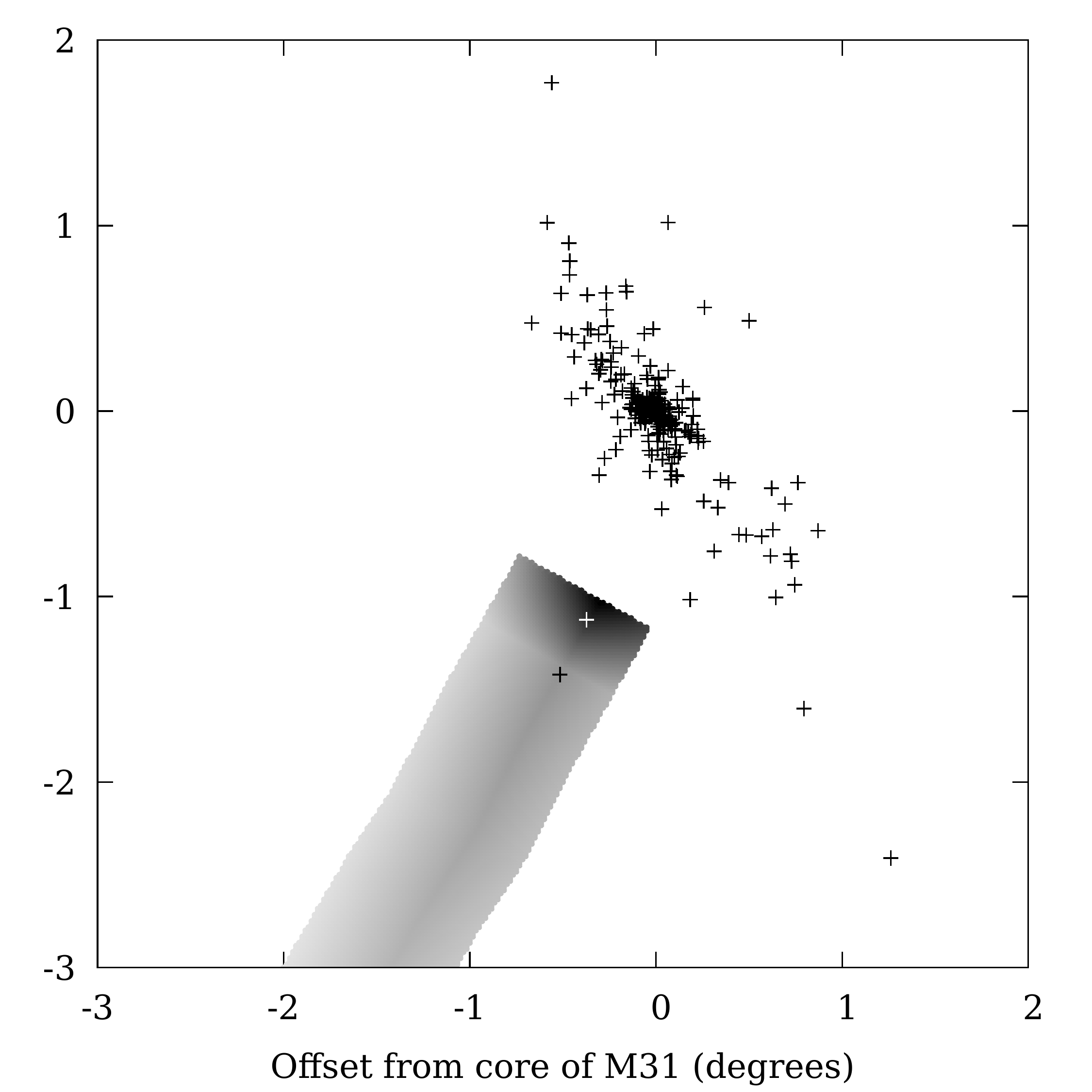}
\caption{
The simplified model of the Stream light used, with the sample of observed novae shown for comparison, cf.\ Figure~\ref{stream_img}.
\label{stream_modellight}}
\end{figure}

At each position along the GSS (i.e.\ roughly perpendicular to the plane of M\,31), and across the GSS, the average stellar light is given by the product of the simple form shown in Figure~\ref{stream_modelapprox}. It can be seen, however, that the component along the southern (negative distance) GSS rises sharply as it nears the plane of M\,31. \citet{2003MNRAS.343.1335M} state that this excess is due to the GSS, and not contamination from the disk light, since it is not mirrored to the north. However, given that we are only concerned with the region where disk novae should be rare, we only consider the GSS more than $1^\circ$ from the plane of M\,31. We also only consider the southern part of the GSS, giving the final GSS light model shown in Figure~\ref{stream_modellight}.

\begin{figure}
\includegraphics[width=\columnwidth]{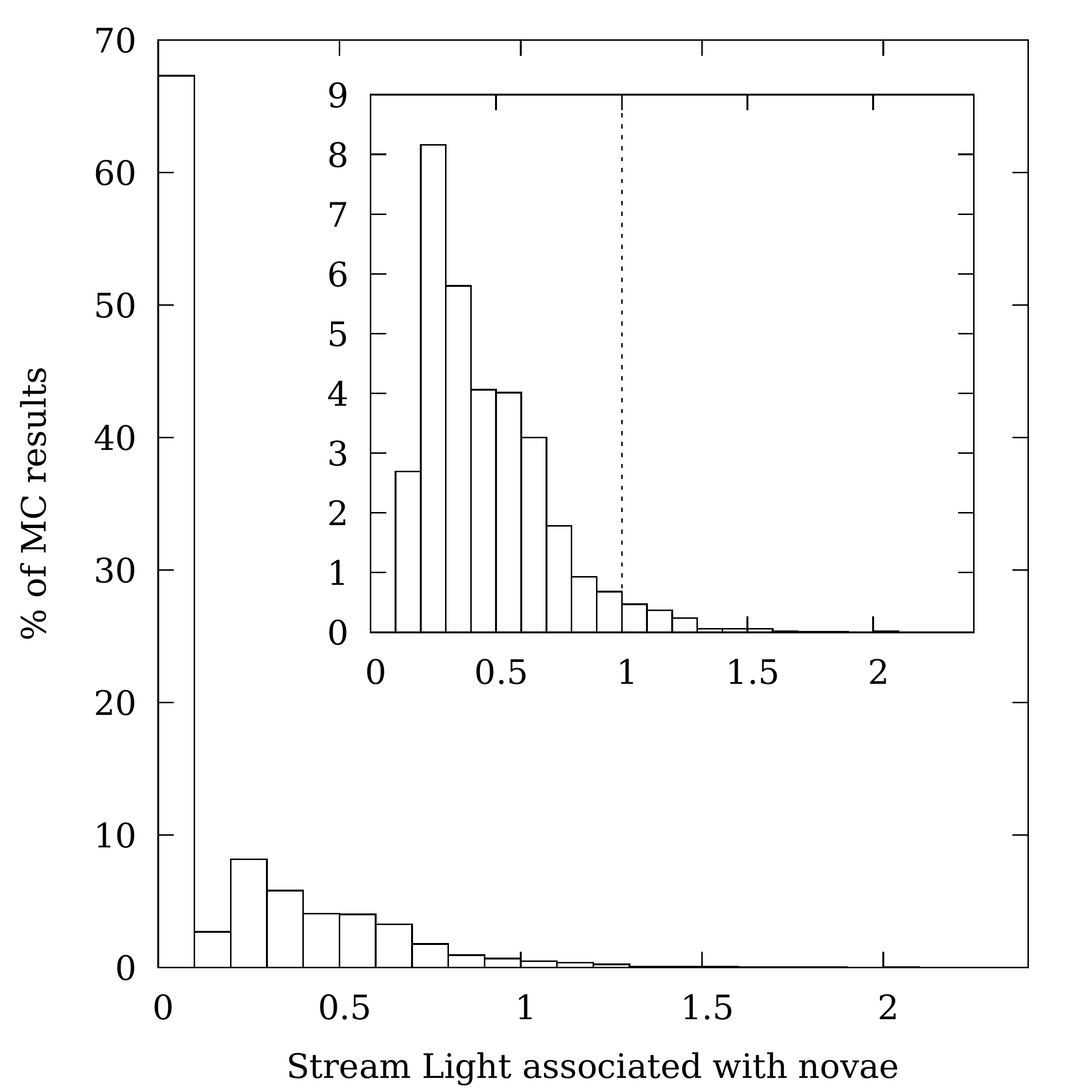}
\caption{
The histogram of GSS light associated with novae in the MC simulations. GSS light is normalised to $1$ for the observed nova sample (i.e.\ AT\,2016dah and AT\,2017fyp). The inset shows the same histogram with the bin at zero GSS light removed. As can be seen, the majority of simulations have no novae on the GSS but there is a tail that stretches past $1$.
\label{mcstream_hist}}
\end{figure}

Given this model, we can determine the association of GSS light with novae (observed or MC simulated) by summing the GSS light at the position of all novae in a given sample. Novae that fall outside the GSS will produce no contribution, while those that fall inside the GSS will contribute in proportion to the stellar light at that point. For the observed sample of novae, only AT\,2016dah and AT\,2017fyp contribute and so we use that value to normalise the result from the MC simulated samples. The result from 10,000 MC simulations of novae distributions is given in the histogram in Figure~\ref{mcstream_hist}. Since the MC simulations only include the disk and bulge, any association with the GSS is random and so this gives an estimate of the likelihood that AT\,2016dah and AT\,2017fyp are {\em coincidentally\/} associated with the GSS.

Although the majority of simulations have no GSS light associated with novae, 1.3\% (132 out of 10,000 simulations) have normalised light greater than $1$ --- in other words they have more GSS light associated with simulated novae than the light associated with the observed nova sample.  As such, we can exclude the hypothesis that the GSS is {\it not} the source of AT\,2016dah and AT\,2017fyp at well beyond the $2\sigma$ confidence limit.

\section{Discussion}\label{sec:discussion}

As an aid to the reader, in Table~\ref{nova_vitals} we provide a summary of key parameters for both AT\,2016dah and 2017fyp.

\begin{table*}
\caption{Summary of key parameters for AT\,2016dah and AT\,2017fyp.\label{nova_vitals}}
\begin{center}
\begin{tabular}{llll}
\hline
Parameter & AT\,2016dah & AT\,2017fyp \\
\hline
R.A.\ (J2000) & $0^\mathrm{h}44^\mathrm{m}41^\mathrm{s}\!\!.05$ & $0^\mathrm{h}45^\mathrm{m}25^\mathrm{s}\!\!.490$ \\
Decl.\ (J2000) & $+40^\circ8^\prime35^{\prime\prime}\!\!.9$ & $+39^\circ50^\prime52^{\prime\prime}\!\!.34$ \\
Discovery date [UT] &2016 Jul 12 & 2017 Aug 7 \\ 
Time of eruption [UT] & 2016 Jul $11.96\pm0.48$ & 2017 Aug $6.08\pm1.48$ \\
\citet*{2010AJ....140...34S} light curve morphology &  Smooth ``S'' & ``S'' or Flat topped ``F'' \\
Time of maximum [days post-eruption] & 2.48 & 4--9\\ 
Peak observed apparent magnitude ($r'$) [mag] & $16.32\pm0.06$ & $17.041\pm0.007$ \\
$r'$-band $t_2$\ /\ $t_3$ decline time [days] & $13.3^{+0.6}_{-0.3}$\ /\ $26\pm2$ & 32--37\ /\ 63--68 \\
$V$-band $t_2$\ /\ $t_3$ decline time [days] & $7\pm1$\ /\ $13\pm1$ & 16--21\ /\ 38--43 \\ 
$B$-band $t_2$\ /\ $t_3$ decline time [days] & $8\pm1$\ /\ $16^{+3}_{-2}$ & 20--25\ /\ 53-58\\
\citet{1964gano.book.....P} speed class ($V$-band) & Very fast & Fast \\
Assumed reddening $E\left(B-V\right)$ [mag] & 0.1 & 0.1\\
\citet{1992AJ....104..725W} spectral taxonomic class & Fe\,{\sc ii}b & Hybrid \\
Radial velocity (heliocentric corrected) [km\,s$^{-1}$] & $-420\pm30$ & $-580\pm50$ \\
H$\alpha$ FWHM [km\,s$^{-1}$] & $2300\pm70$ & $2550\pm60$ \\ 
\hline
\end{tabular}
\end{center}
\end{table*}

\subsection{AT 2016dah}

With a $V$-band $t_2=7\pm1$\,d and exhibiting expansion velocities (inferred from the Balmer emission line FWHM and P\,Cygni terminal velocities) of 2300\,km\,s$^{-1}$, this rapid evolution of AT\,2016dah is at the extremes of those observed for Fe\,{\sc ii} nova in M\,31. In their photometric and spectroscopic survey of 48 new M\,31 novae (91 in total), the fastest light curve evolution reported by  \citet{2011ApJ...734...12S} for an Fe\,{\sc ii} nova was $t_2(V)=8.9\pm0.2$\,d for the luminous nova M31N\,2009-10b. However, the H$\alpha$ line width of M31N\,2009-10b only reached $\sim1700$\,km\,s$^{-1}$ \citep{2009ATel.2251....1B,2009ATel.2248....1D}. Indeed, none of the Fe\,{\sc ii} novae reported by \citet[see their Figure~16]{2011ApJ...734...12S} displayed H$\alpha$ FWHMa $>2000$\,km\,s$^{-1}$. Hybrid novae, those that simultaneously show elements of the He/N and Fe\,{\sc ii} spectral classes, or evolve from one to another, are also referred to as Fe\,{\sc ii}b or `broad' novae \citep[see, for e.g.,][]{1998ApJ...506..818D}. Similarly to AT\,2016dah, M31N\,2005-09b displayed Fe\,{\sc ii}, Na\,D, and He\,{\sc i} emission, although the FWHM  was only $\approx2000$\,km\,s$^{-1}$, \citet{2011ApJ...734...12S} classified this system as a Fe\,{\sc ii}b (when including the hybrid M31N\,2006-10b, just 2 of 91 spectroscopically confirmed novae in M\,31 at that time were hybrids or Fe\,{\sc ii}b novae). If we turn to M\,33 for comparison, similarly to their M\,31 work, \citet{2012ApJ...752..156S} reported photometric and spectroscopic observations of eight novae in M\,33. Three of those novae, M33N\,2003-09a, 2010-10a, and 2010-11a were classified as Fe\,{\sc ii}b, with H$\alpha$ FWHM velocities of 2700, 4210, and 2610\,km\,s$^{-1}$, respectively. Here, M33N\,2003-09a and 2010-11a, which show similar velocities to AT\,2016dah, displayed `typical' Fe\,{\sc ii} spectra, where 2010-10a (like AT\,2017fyp) present elements of both Fe\,{\sc ii} and He/N spectra -- a hybrid. It should be noted that the spectral properties of the vast majority of the \citet{2011ApJ...734...12S,2012ApJ...752..156S} novae were derived from a single snap-shot spectrum taken at essentially random times during their early evolution. Without detailed spectral sequences, as were obtained for both AT\,2016dah and AT\,2017fyp, it is always possible that a single spectrum does not reveal the whole picture of the evolution of a given nova (particularly if the contextual information provided by a well sampled light curve is unavailable).

The Heliocentric recession velocity of AT\,2016dah is $-420\pm30$\,km\,s$^{-1}$, this is formally inconsistent (at $4\sigma$) with the accepted Heliocentric recession velocity of M\,31, $-300\pm4$\,km\,s$^{-1}$ \citep{1991rc3..book.....D}. We will discuss the interpretation of this apparent discrepancy in Section~\ref{sec:stream}.

\citet{2016ATel.9382....1M} reported 15\,GHz radio observations of AT\,2016dah using the Arcminute Microkelvin Imager (AMI) Large Array $\sim25$\,d post-eruption. Those authors reported a $3\sigma$ upper limit of 102\,$\mu$Jy or a spectral luminosity $<7.2\times10^{22}$\,erg\,s$^{-1}$\,Hz.

\subsection{AT 2017fyp}

AT\,2017fyp displays elements of both Fe\,{\sc ii} and He/N spectra simultaneously throughout its early decline. There is no evidence, over the spectral time-series collected of a transition from one spectral type to the other, although it is possible that Fe\,{\sc ii} would have been more dominant should earlier spectra have been available. Given the spectral development and a mean FWHM of $2550\pm60$\,km\,s$^{-1}$, AT\,2017fyp is typical of a hybrid spectral type. As with the similar Fe\,{\sc ii}b systems, hybrids appear rare in the M\,31 population \citep[$\sim2$\%;][]{2011ApJ...734...12S}, but appear possibly more common in younger stellar populations such as M\,33 \citep{2012ApJ...752..156S}.

The light curve of AT\,2017fyp reveals a short lived, $\sim5$\,d, plateau. As such, we chose to classify AT\,2017fyp as an `F'-type or flat-topped nova \citep[after][]{2010AJ....140...34S}. The definition of such novae from \citet{2010AJ....140...34S} describes a flat-top that lasts 2--8 months, much longer than seen for AT\,2017fyp, those authors also indicate that only a handful of Galactic novae are F-type. However, \citet{2010AJ....140...34S} suggest that V2295\,Ophiuchi, which exhibited a flat-top lasting $\sim8$\,d, may also fall into this class. The physical mechanism driving the flat-top phenomenon is still unclear.

AT\,2017fyp shows an even greater discrepancy with the Heliocentric recession velocity of M\,31 than AT\,2016dah. The recession velocity of $-580\pm50$\,km\,s$^{-1}$ differs from that of M\,31 beyond $5\sigma$. Again, this shall be discussed in Section~\ref{sec:stream}.

\subsection{Maximum magnitude---rate of decline}

The maximum magnitude---rate of decline \citep[MMRD;][]{1945PASP...57...69M} relationship has been employed for some time to estimate distances to Galactic and extragalactic novae. In recent years the validity of that relationship has been called into question by a number of authors (see, for e.g., \citealt{2017ApJ...839..109S} and \citealt{2018MNRAS.481.3033S}). However, \citet{2019A&A...622A.186S} and, more recently, \citet[who performed a reanalysis of existing MMRD data]{2020arXiv200406540D} refute such claims. In either case, due to the inherent scatter, the MMRD is not a reliable distance indicator to \textit{individual} novae (such as the subjects of this paper) and is strongest when used to estimate a distance toward a population of novae (see \citealt{2019arXiv190910497D} and \citealt{2020arXiv200406540D} for discussions regarding recurrent novae). The MMRD does not have the sensitivity to discriminate between \textit{individual} novae in the GSS and those within M\,31.

Here, we simply employ the MMRD to explore whether the luminosity and decline times of AT\,2016dah and AT\,2017fyp are typical of the  M\,31 population. \citet{2006MNRAS.369..257D} derived the only $r'$-band MMRD for M\,31 novae (see their Equation~3). Using the measured $r'$-band $t_2$ times for AT\,2016dah and AT\,2017fyp, their MMRD would predict peak apparent magnitudes of $r'=16.2$ and 16.8, respectively. Both these MMRD predictions agree well with our observations. Additionally, the MMRD indicates that both novae are indeed of a typical luminosity for their speed class, when compared with the global M\,31 population.

\subsection{X-ray emission (or lack thereof)}

AT\,2016dah was observed with the XRT onboard {\it Swift} at eleven epochs between $7\leq\Delta t\leq87$\,d post-eruption. AT\,2017fyp was observed with the XRT at ten epochs between $19\leq\Delta t\leq171$\,d post-eruption. No X-ray photons were detected for either source, see Table~\ref{swift_data}. At the distance of (and column toward) M\,31, the super-soft X-ray source (SSS) of nova eruptions are regularly detected \citep[see, for e.g.,][]{2010A&A...523A..89H,2011A&A...533A..52H,2014A&A...563A...2H}, although a good proportion have gone undetected despite available X-ray observations. \citet{2014A&A...563A...2H} observed correlations between optical and X-ray parameters for M\,31 novae. These included the unveiling of the SSS ($t_\mathrm{on}$) with $t_2$ and the expansion velocity $v_\mathrm{exp}$, as estimated from optical spectra, and between the end of the SSS phase ($t_\mathrm{off}$) with $t_\mathrm{on}$.

Utilising Equation~6 from \citet{2014A&A...563A...2H}, we estimate the epoch of $t_\mathrm{on}$ using our $r'$-band estimates of $t_2$ to be:\ $t_\mathrm{on}=70\pm20$\,d and $150\pm70$\,d post-eruption for AT\,2016dah, and AT\,2017fyp, respectively. Based on the H$\alpha$ FWHM measurements, which is assumed to be representative of the expansion velocity, we estimate \citep[using Equation~7 from][]{2014A&A...563A...2H}:\ $t_\mathrm{on}\lesssim90$\,d and $t_\mathrm{on}\lesssim80$\,d post-eruption for AT\,2016dah, and AT\,2017fyp, respectively. The estimates from both methods for AT\,2016dah are consistent, but given the larger expansion velocity yet slower decline of AT\,2017fyp those estimates differ notably. We similarly estimate $t_\mathrm{off}$ using Equation~4 from \citet{2014A&A...563A...2H} to be in the range:\ $100\lesssim t_\mathrm{off}\lesssim350$\,d for AT\,2016dah, and $150\lesssim t_\mathrm{off}\lesssim740$\,d or $t_\mathrm{off}\lesssim260$\,d for AT\,2017fyp for the $t_2$ and $v_\mathrm{exp}$ methods, respectively. 

Additionally, the appearance of He\,{\sc ii} 4686\,\AA\ in a nova spectrum can be an indication that part of the ejecta, or any surviving, or reformed, accretion structure is being directly illuminated by the SSS. As such, the appearance of He\,{\sc ii} often accompanies the unveiling of the SSS. For AT\,2016dah, He\,{\sc ii} may have appeared as early as day 14 (but this could also be O\,{\sc ii} emission), by day 41 the identification of He\,{\sc ii} is clearer. For AT\,2017fyp, He\,{\sc ii} emission appears at day 55 post-eruption. Both these epochs are consistent with the estimates using the \citet{2014A&A...563A...2H} relations.

Therefore is seems clear that there is a good likelihood that the {\it Swift} observations did sample the SSS phase of both AT\,2016dah and AT\,2017fyp. However, the SSS luminosity of these systems was below the detection limits of those observations, implying SSS luminosities $L_\mathrm{SSS}\lesssim2\times10^{36}$\,erg\,s$^{-1}$ (sampled between 0.3--10\,keV). Other possibilities include the ejecta only becoming optically thin to X-rays {\it after} the SSSs had turned-off (i.e.\ formally $t_\mathrm{on}>t_\mathrm{off}$). Or that the SSSs turned-on after both {\it Swift} campaigns had ended, given the behaviour of the \cite{2014A&A...563A...2H} M\,31 sample, this seems unlikely for such fast novae. Finally, $t_\mathrm{off}$ may have occurred before the first {\it Swift} observations, i.e.\ before day 7 and 18 post-eruption for AT\,2016dah and AT\,2017fyp, respectively. However, only two novae have exhibited SSS phases so short, the recurrent novae V745\,Scorpii \cite[$t_\mathrm{off}\sim6$\,days;][]{2015MNRAS.454.3108P} and M31N\,2008-12a \citep[$t_\mathrm{off}\sim15$\,days, but only for the peculiarly late and short 2016 eruption;][]{2018ApJ...857...68H}. Such short SSS phases are due to a TNR on the surface of a near-Chandrasekhar mass WD, the critical/ignition mass is low and the surface conditions particularly extreme. However, neither AT\,2016dah or AT\,2017fyp show any properties that would imply that they may be recurrent novae \citep[for e.g., extremely fast light curve evolution, under-luminous eruptions, high ejection velocities, (post-maximum) light curve plateaus, ejecta deceleration, evolved donor, etc.; see the discussions within][]{2014ApJ...788..164P,2019arXiv191213209D}.

\subsection{Quiescent systems}

\citet{2012ApJ...746...61D} demonstrated that the evolutionary state of the donor in a nova system could be determined by the position of that quiescent nova on a colour-magnitude diagram. This technique is particularly sensitive for those systems with giant donors, where the donor dominates the emission from the optical through to the mid/far-IR \citep[also see][]{2014MNRAS.444.1683E} at quiescence. \citet{2014ASPC..490...49D} demonstrated that at the distance of M\,31 and M\,33 all quiescent novae with red giant donors could be recovered, with sufficiently deep and high spatial resolution imaging. \citet{2016ApJ...817..143W} and particularly \citet{2017ApJ...849...96D} went on to show that the accretion disk of systems with high mass transfer rates were recoverable in the Local Group by utilising the near-UV.

\begin{figure*}
    \begin{minipage}{0.9\textwidth}
    \includegraphics[width=0.33\textwidth]{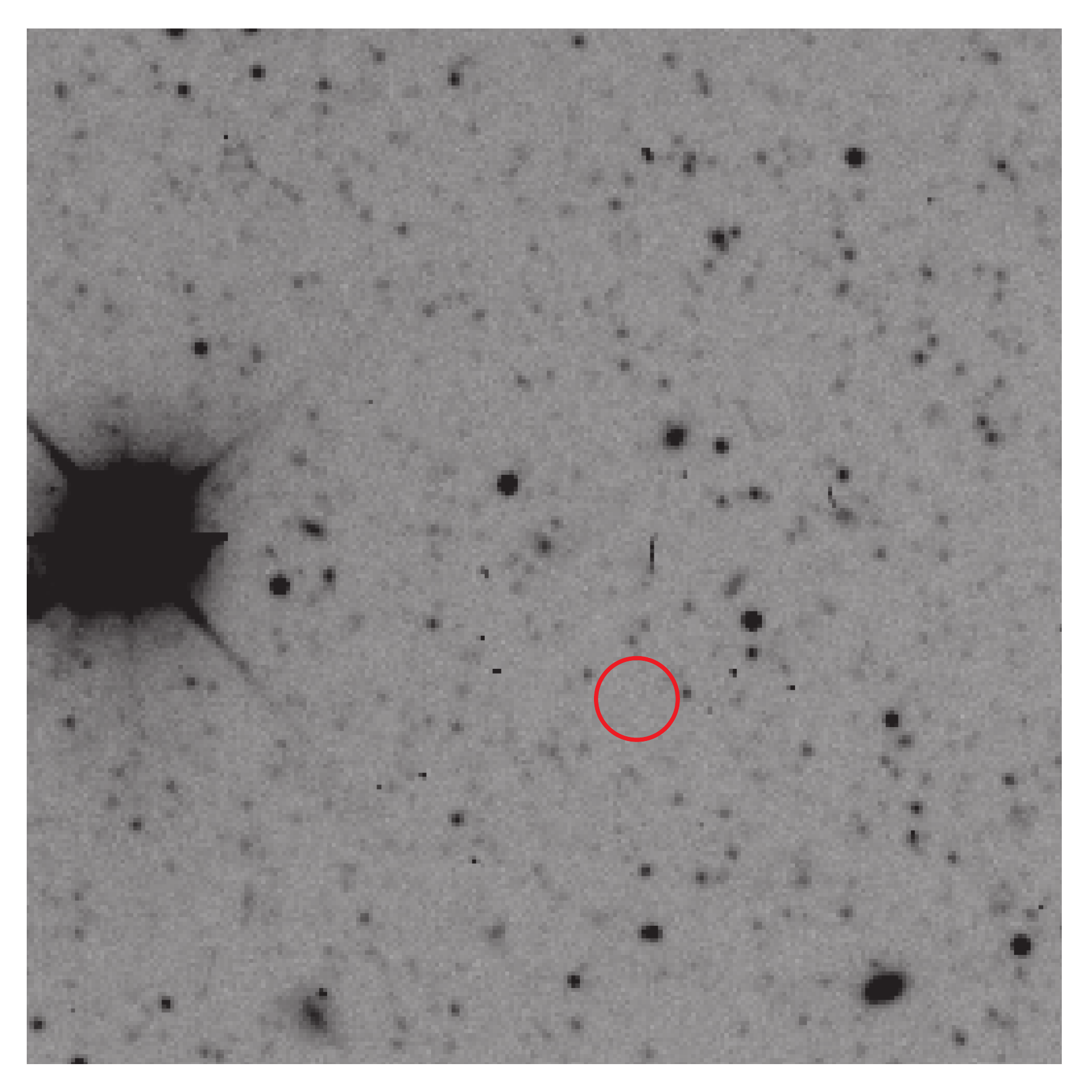}\hfill
    \includegraphics[width=0.33\textwidth]{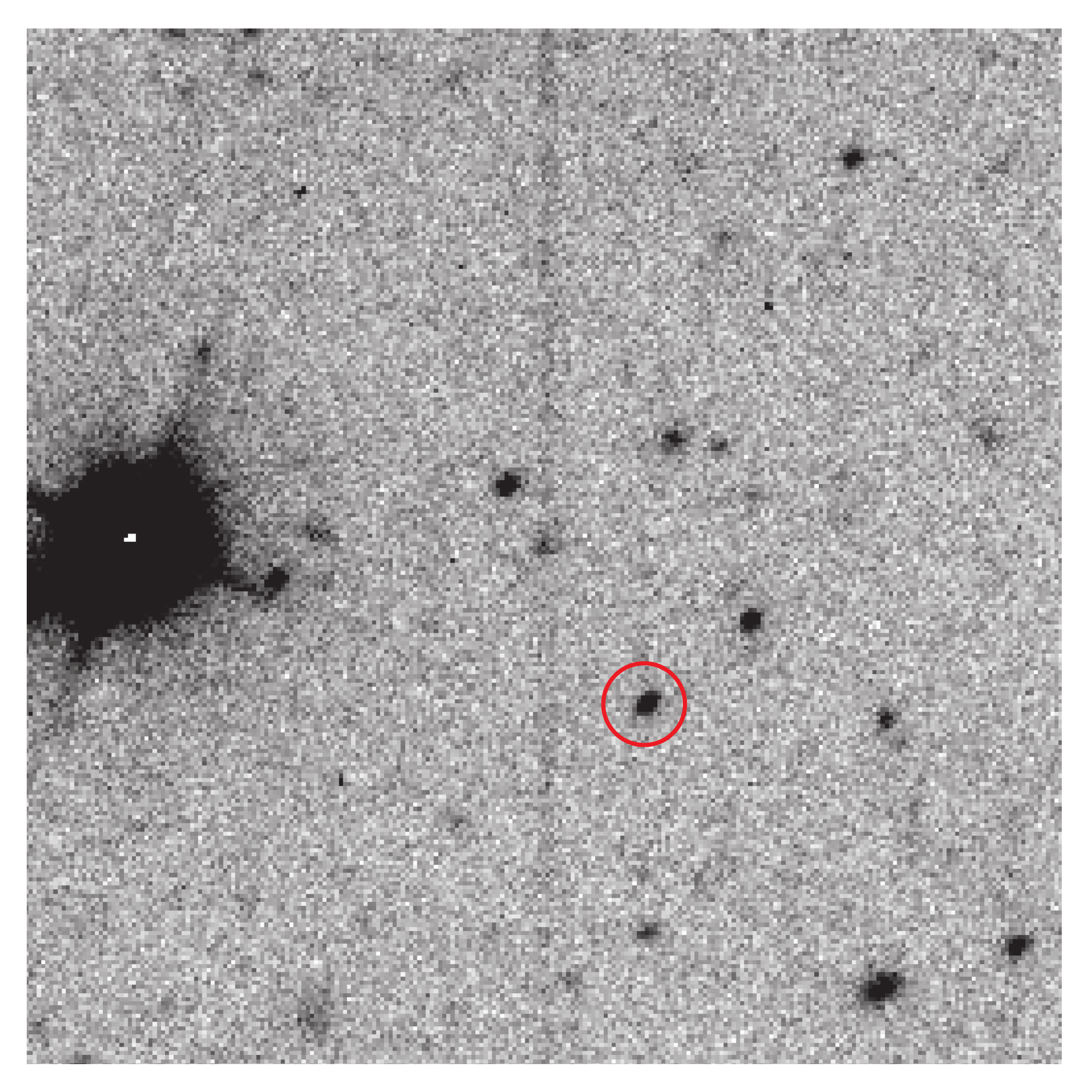}\hfill
    \includegraphics[width=0.33\textwidth]{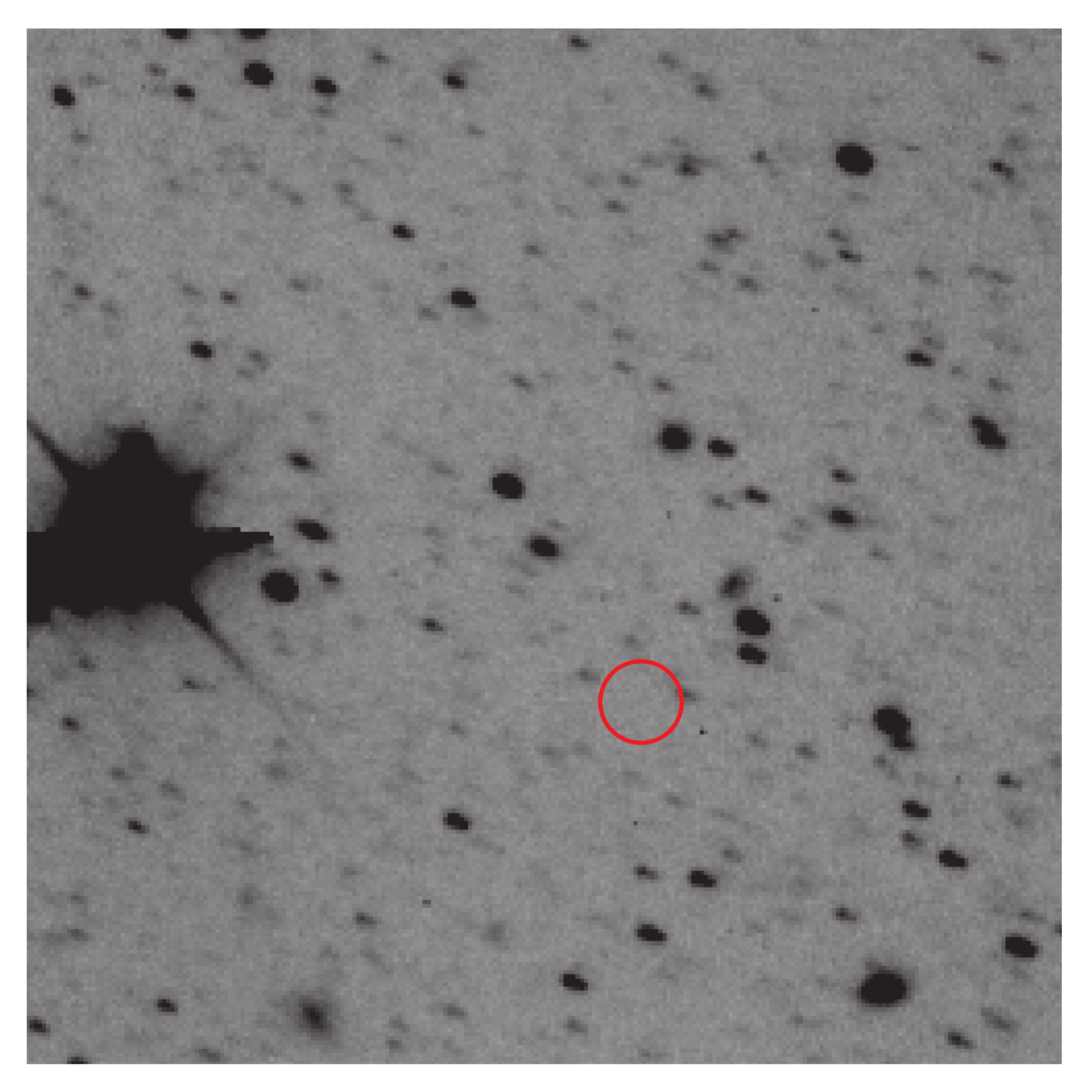}
    \\
    \includegraphics[width=0.33\textwidth]{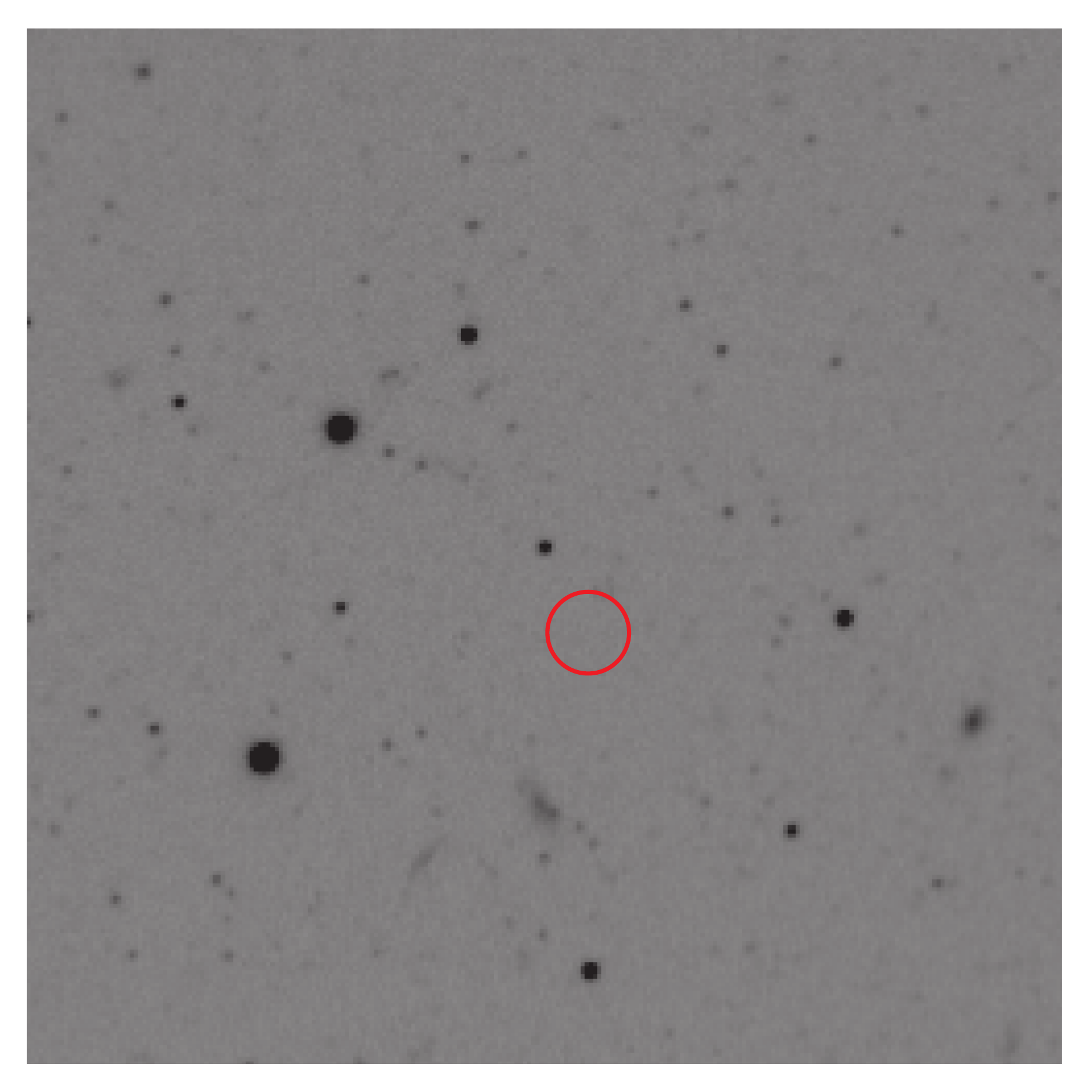}\hfill
    \includegraphics[width=0.33\textwidth]{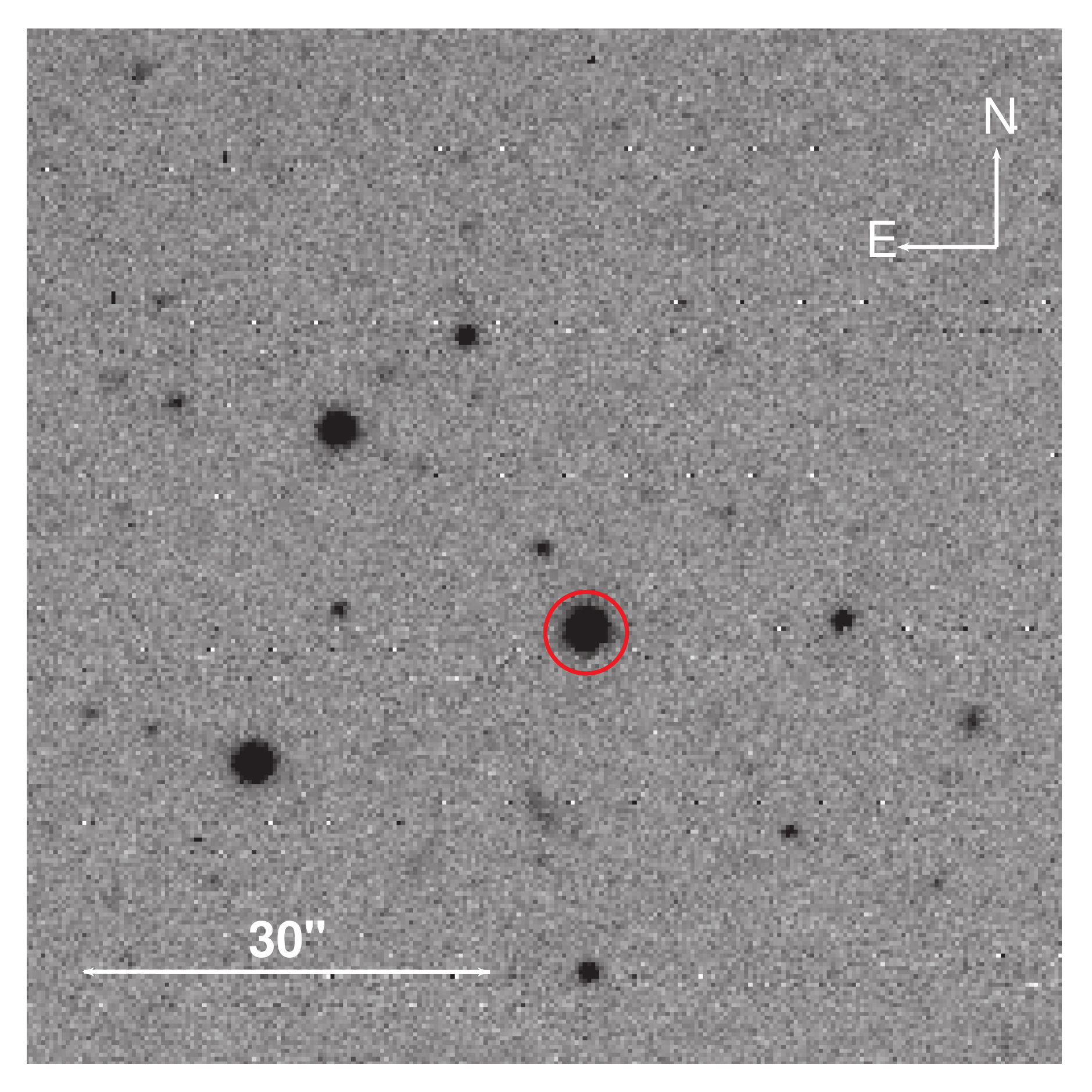}\hfill
    \includegraphics[width=0.33\textwidth]{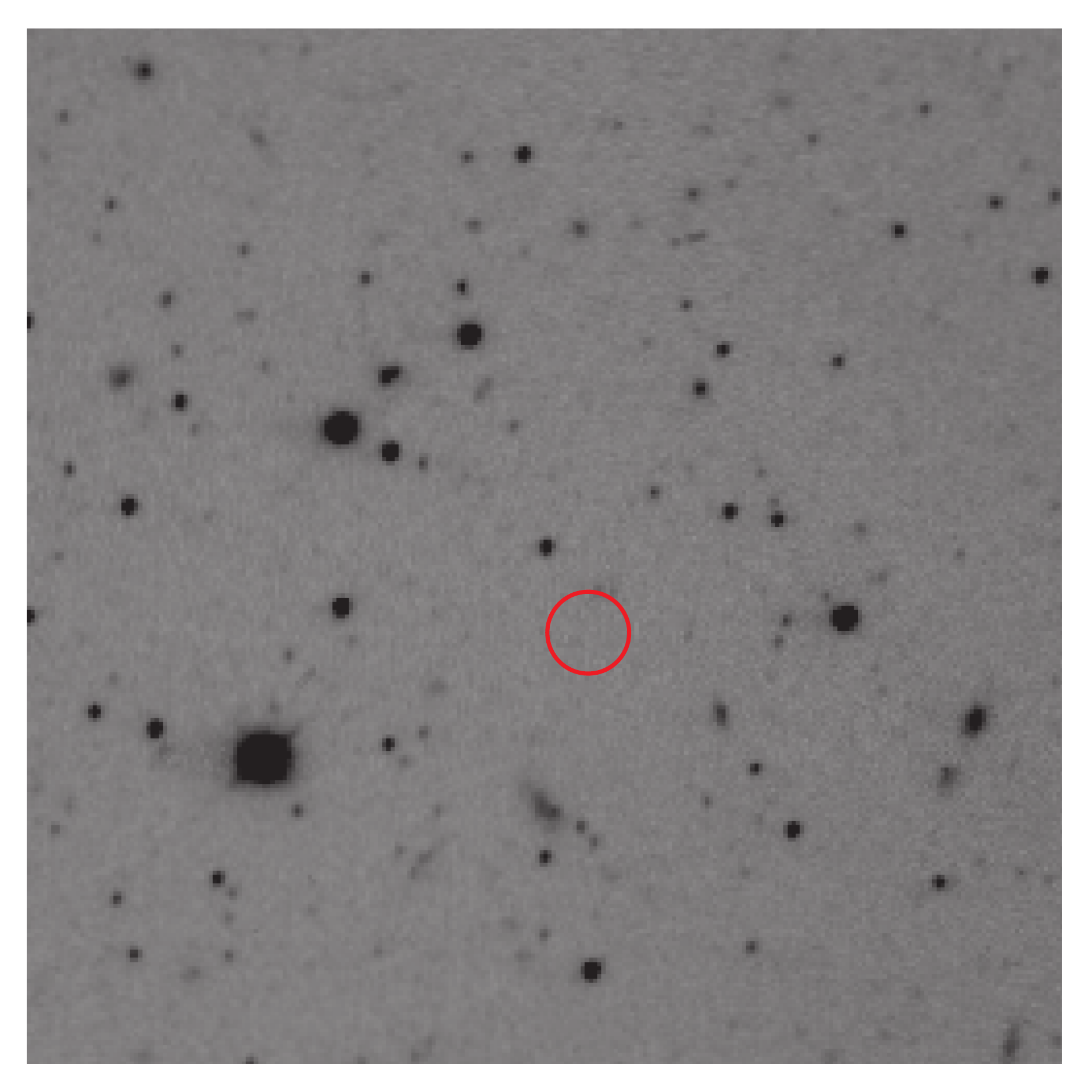}
    \end{minipage}
    \caption{Progenitor search regions for AT\,2016dah (top row) and AT\,2017fyp (bottom row). The central image pair show Liverpool Telescope $r'$-band eruption images ($3\times180$\,s exposure) with the erupting nova indicated by the red circle (radius $3^{\prime\prime}$). The images to the left and right show regions of the CFHT CFH12K mosaic data ($3\times545$\,s) around each nova \citep{2003MNRAS.343.1335M}, the nova position is again centred in the red circle; the left hand-image is $V$-band data, the right is $I$-band. All images are approximately a square arc-minute and have the same orientation, the image scale and orientation are indicated in the bottom-centre image. The CFHT data were not of sufficient depth to recover the progenitors systems (see text for details).}
    \label{fig:quiescent}
\end{figure*}

In Figure~\ref{fig:quiescent} we show the regions around AT\,2016dah and AT\,2017fyp in the \citet{2003MNRAS.343.1335M} CFHT data. As is shown in these images, there are no detected sources within at least $1^{\prime\prime}$ of either nova. Photometry at the position of the nova in the CFHT data gives the following quiescent upper limits:\ AT\,2016dah, $V>24.8$\,mag and $I>24.3$\,mag; AT\,2017fyp, $V>24.5$\,mag and $I>22.4$\,mag. Although they provide the best UV coverage of location of the two novae, the GALEX data are not of sufficient depth or spatial resolution to be useful. At the distance of M\,31, the Galactic red giant donor RNe V3890\,Sagittarii and T\,Coronae Borealis would have $I$-band magnitudes of $\sim21.5$ and $\sim22.2$, respectively \citep{2012ApJ...746...61D}. The depth of the CFHT data are, however, sufficient to rule out such luminous red giant donors for both these systems. Therefore it is most likely that the mass donor in both systems is a main sequence star, although we also cannot rule out sub-giant donors for either system.

\subsection{Giant Stellar Stream novae}\label{sec:stream}

Taken at face value, Figure~\ref{stream_img}, which shows the location of AT\,2016dah and AT\,2017fyp with respect to the GSS, is extremely suggestive that both novae should be strongly associated with the GSS. But upon further inspection, one does notice that there are a small number of other spectroscopically confirmed novae that appear beyond the bulk of the {\it typical} M\,31 bulge--disk novae. Our analysis of the M\,31 nova spatial distribution (see Section~\ref{sec:spatial}) does not definitively confirm that both AT\,2016dah and AT\,2017fyp are associated with the GSS. But it does indicate that the likelihood of, at least, two M\,31 disk or bulge novae being spatially associated with the GSS by chance, at least as strongly as these two, is small, $\sim1\%$.

Both AT\,2016dah and AT\,2017fyp reside within Field~7 of the CFHT targeted survey of the GSS undertaken by \citet{2003MNRAS.343.1335M}. A radial velocity survey of the stream was reported by \citet{2004MNRAS.351..117I}. That survey targeted four of the thirteen \citet{2003MNRAS.343.1335M} fields, unfortunately it did not cover field 7, but did survey fields 6 and 8, located either side of field 7. \citet{2004MNRAS.351..117I} reported a strong velocity gradient along the stream. The southern most tip has similar Heliocentric radial velocity to M\,31 ($\sim-300$\,km\,s$^{-1}$), whereas where the stream appears to coincide with M\,31 it is approaching us relative to M\,31 ($\sim-600$\,km\,s$^{-1}$). The radial velocity within field 6 was $\sim-480$\,km\,s$^{-1}$. As such, given the trend of radial velocity along the stream, we would expect the radial velocity of stream stars within field 7 to lie between -480 and -600\,km\,s$^{-1}$. The radial velocity of AT\,2016dah and AT\,2017fyp are $-420\pm30$ and $-580\pm50$\,km\,s$^{-1}$, respectively. Our radial velocity measurements will probably suffer from additional systematic uncertainties from, for e.g., the nova ejecta geometry, ejecta self-absorption, the low spectral resolution, and possibly even the binary orbital motion. Even so, these radial velocity measurements are strongly suggestive that both novae are members of the GSS.

Neither the spatial distribution analysis nor the radial velocity result alone provide the proverbial {\it smoking gun}. But, taken together they present compelling evidence for both AT\,2016dah and AT\,2017fyp to be considered members of the GSS and therefore not M\,31 novae. As such, we will proceed under the assumption that both nova {\it are} associated with the GSS. But, with just two examples, it is hard to draw any concrete conclusions, nevertheless, one can speculate.

It is likely that the bulk of the stars associated with the GSS arose from the hitherto unidentified progenitor galaxy \citep[see, e.g.,][and references therein]{2013MNRAS.434.2779F,2017MNRAS.464.3509K,2018MNRAS.475.2754H}. Therefore, it would seem probable that both AT\,2016dah and AT\,2017fyp were originally also part of the GSS progenitor. Recent simulations of the GSS formation have predicted that the progenitor was a small spiral galaxy  \citep[see, e.g.,][]{2013MNRAS.434.2779F,2017MNRAS.464.3509K}, perhaps not too unlike the present day M\,33, although with a mass more akin to the Large Magellanic Cloud. As previously mentioned, \citet{2012ApJ...752..156S} found significant differences between the spectroscopic properties of M\,31 and M\,33 novae. Fe\,{\sc ii}b and hybrid novae, such as AT\,2016dah and AT\,2017fyp, respectively, are rare among the M\,31 population \citep[$\sim2\%$;][]{2011ApJ...734...12S}; whereas such novae make up around half of those in M\,33 (albeit it based on a sample of eight). However, the stellar mass of the GSS is substantially lower than M\,33 or the LMC \citep[$\sim2.4\times10^8$\,M$_\odot$;][]{2001Natur.412...49I,2006MNRAS.366.1012F}, similar to that of the Local Group dwarf irregular NGC\,6822 \citep*{2003MNRAS.340...12W}. Based on the `Luminosity Specific Nova Rate' \citep[see][]{2014ASPC..490...77S,2019arXiv190910497D}, the nova rate from such a stellar mass would be expected to be small:\ $\sim0.1$\,yr$^{-1}$.

We should therefore ask why have two GSS novae been discovered only a year apart? In part the answer to this lies in the history of nova (and all variable and transient) surveys in and around M\,31. As discussed in Section~\ref{sec:spatial}, the completeness of the M\,31 nova catalogue, particularly the spectroscopically confirmed catalogue, has both spatial and temporal dependence. M\,31 nova survey strategy has (almost) always focussed on obtaining the largest amount of novae possible. In the early days, this meant focussing small fields on the bulge. As detector technology evolved, surveys expanded to cover more and more of the disk -- but always with diminishing returns when considering the total nova yield \citep[see][for a fuller discussion]{2019arXiv190910497D}. In the past decade or so, high-cadence large-fields surveys -- such as those employed here:\ ASAS-SN and PTF/iPTF/ZTF -- have been able to probe the large volume around M\,31, indeed the majority of the sky. So it is probably fair to say that we have only been capable of discovering (and confirming) GSS novae for around a decade. Given a nova rate of $\sim0.1$\,yr$^{-1}$, a yield of two novae in that time would be well within expectations. As the temporal baseline of the GSS (indeed all such streams in the Local Group) grows over the coming years and decades, we would expect to detect more such novae, and hopefully address some of the questions raised here.

\section{Conclusions}\label{sec:conclusions}

In this Section we summarise the main findings of this paper.

\begin{enumerate}
\item[1)] The classical novae AT\,2016dah and AT\,2017fyp are located far to the south of the bulk of the content of M\,31 and that host's nova population.
\item[2)] AT\,2016dah was discovered and followed photometrically and spectroscopically well before reaching maximum light. The initial spectra probe the fireball phase where the emission is dominated by a black body-like continuum.
\item[3)] AT\,2016dah is a very fast nova with a S-type light curve that displays prominent Fe\,{\sc ii} emission lines during its early decline. High ejecta velocities lead to a classification as a (rare for M\,31) Fe\,{\sc ii}b nova. 
\item[4)] AT\,2017fyp was a fast nova, possibly with an F-type light curve, whose early decline spectra simultaneously contained Fe\,{\sc ii} and He/N lines, leading to a classification as a hybrid nova.
\item[5)] Both novae were followed well into their nebular phase, in part aided by the low surface brightness so far from the centre of M\,31.
\item[6)] Despite reasonable sampling by the Neil Gehrels {\it Swift} Observatory, the super-soft X-ray source of neither nova was detected. We propose that this is most likely due to the X-ray emission being below the detection limit of those observations.
\item[7)] A progenitor search within available archival data revealed no detected quiescent counterpart for either nova. We can rule out luminous red giant donors, cf.\ T\,CrB, and, as such, we suggest that both systems are most likely to harbour main sequence donors.
\item[8)] Hybrid and Fe\,{\sc ii}b novae are rare within the M\,31 population, perviously accounting for just 2\% of spectroscopically confirmed novae.
\end{enumerate}

Both  AT\,2016dah and AT\,2017fyp appear strongly associated with the Giant Stellar Stream to the south of M\,31. The radial velocities of both novae imply an association with the GSS. The distribution of novae away from the bulge are elongated along the major axis of the inclined disk of M\,31 and therefore are inconsistent with the broadly spherically-symmetric halo. However, our Monte Carlo simulations of the M\,31 bulge and disk nova populations allows us to rule out a chance alignment of these two novae with the GSS at well beyond $2\sigma$. Combined, this evidence leads us to claim that both novae are associated with the GSS, indeed they are the first to be associated with any tidal stellar stream. Therefore, it would seem probable that these nova systems formed within the GSS progenitor galaxy and are therefore not associated with the M\,31 nova population. 
 
\section*{Acknowledgements}

We would like to express our gratitude to Massimo Della Valle for his helpful and thoughtful comments when refereeing the original manuscript. 
The authors would like to thank Conor Ransome (and collaborators) for advanced access to the extended M\,31 nova spectroscopic catalogue.
MJD would like to thank Kim Page for guidance with respect to reduction and analysis of {\it Swift} UVOT and XRT data, and Andreea Font for discussion regarding the Giant Stellar Stream. 
MJD and AMN also acknowledge funding from the UK Science and Technology Facilities Council (STFC) consolidated grant ST/R000484/1. 
ALJ acknowledges funding from INTO Newcastle University. 
IDWH acknowledges funding from Student Finance England. 
This work was supported in part by the GROWTH (Global Relay of Observatories Watching Transients Happen) project funded by the National Science Foundation Partnership in International Research and Education program under Grant No 1545949. GROWTH is a collaborative project between California Institute of Technology (USA), Pomona College (USA), San Diego State University (USA), Los Alamos National Laboratory (USA), University of Maryland College Park (USA), University of Wisconsin Milwaukee (USA), University of Washington Seattle (USA), Texas Tech University (USA), Tokyo Institute of Technology (Japan),  National Central University (Taiwan), Indian Institute of Astrophysics (India), Inter-University Center for Astronomy and Astrophysics (India), Weizmann Institute of Science (Israel), The Oskar Klein Centre at Stockholm University (Sweden), Humboldt University (Germany), Liverpool John Moores University (LJMU; UK), University of Sydney (Australia).
The Liverpool Telescope is operated on the island of La Palma by LJMU in the Spanish Observatorio del Roque de los Muchachos of the Instituto de Astrof\'{i}sica de Canarias with financial support from STFC. 
IRAF is distributed by the National Optical Astronomy Observatories, which are operated by the Association of Universities for Research in Astronomy, Inc., under cooperative agreement with the National Science Foundation.
This research has made use of data and software provided by the High Energy Astrophysics Science Archive Research Center (HEASARC), which is a service of the Astrophysics Science Division at NASA/GSFC.
Funding for SDSS-III has been provided by the Alfred P. Sloan Foundation, the Participating Institutions, the National Science Foundation, and the U.S. Department of Energy Office of Science. The SDSS-III web site is \url{http://www.sdss3.org}. SDSS-III is managed by the Astrophysical Research Consortium for the Participating Institutions of the SDSS-III Collaboration including the University of Arizona, the Brazilian Participation Group, Brookhaven National Laboratory, Carnegie Mellon University, University of Florida, the French Participation Group, the German Participation Group, Harvard University, the Instituto de Astrofisica de Canarias, the Michigan State/Notre Dame/JINA Participation Group, Johns Hopkins University, Lawrence Berkeley National Laboratory, Max Planck Institute for Astrophysics, Max Planck Institute for Extraterrestrial Physics, New Mexico State University, New York University, Ohio State University, Pennsylvania State University, University of Portsmouth, Princeton University, the Spanish Participation Group, University of Tokyo, University of Utah, Vanderbilt University, University of Virginia, University of Washington, and Yale University.
This work has made use of data from the European Space Agency (ESA) mission
{\it Gaia} (\url{https://www.cosmos.esa.int/gaia}), processed by the {\it Gaia}
Data Processing and Analysis Consortium (DPAC,
\url{https://www.cosmos.esa.int/web/gaia/dpac/consortium}). Funding for the DPAC
has been provided by national institutions, in particular the institutions
participating in the {\it Gaia} Multilateral Agreement.

\bibliographystyle{mnras}
\bibliography{refs} 
\label{lastpage}

\clearpage

\begin{appendix}

\section{Standards \& Photometry --- To appear on-line only}

\begin{landscape}
\begin{table}
\caption{Pan-STARRS sources used for calibration of the AT\,2016dah photometry.}\label{16full_standards} 
\begin{center}
\begin{tabular}{lllllllll}
\hline
\# & PANSTARRS ID & R.A.\ [J2000] & Dec.\ [J2000] & $u'$ [mag] & $g$ [mag] & $r$ [mag] & $i$ [mag] & $z$ [mag] \\
\hline
001 & 156090110644167056 & 00:44:15.463 & 40:04:49.243 & $17.656\pm0.009$ & $16.3037\pm0.0033$ & $15.8897\pm0.0022$ & $15.7505\pm0.0029$ & $15.7308\pm0.0063$ \\ 
002 & 156160110656529961 & 00:44:15.760 & 40:08:27.989 & \ldots & $17.9499\pm0.0054$ & $17.0009\pm0.0028$ & $16.5633\pm0.0036$ & $16.3556\pm0.0055$ \\ 
003 & 156130110677692991 & 00:44:16.260 & 40:06:37.018 & \ldots & $17.7762\pm0.0032$ & $16.5773\pm0.0045$ & $15.7959\pm0.0033$ & $15.4573\pm0.0023$ \\ 
004 & 156200110699024800 & 00:44:16.773 & 40:10:12.546 & $18.554\pm0.015$ & $17.3553\pm0.0038$ & $16.9734\pm0.0040$ & $16.8323\pm0.0019$ & $16.8126\pm0.0062$ \\ 
005 & 156130110738743484 & 00:44:17.733 & 40:06:38.571 & \ldots & $17.2886\pm0.0038$ & $16.7190\pm0.0024$ & $16.4995\pm0.0024$ & $16.4233\pm0.0046$ \\ 
006 & 156200110751752256 & 00:44:18.039 & 40:10:04.912 & $18.353\pm0.013$ & $17.1149\pm0.0042$ & $16.7029\pm0.0021$ & $16.5488\pm0.0023$ & $16.5033\pm0.0050$ \\ 
007 & 156210110791621242 & 00:44:18.996 & 40:10:31.895 & \ldots & $17.4554\pm0.0046$ & $16.7802\pm0.0044$ & $16.5194\pm0.0023$ & $16.4190\pm0.0027$ \\
\hline
\end{tabular}
\end{center}
\end{table}
\end{landscape}

\clearpage

\begin{landscape}
\begin{table}
\contcaption{Pan-STARRS sources used for calibration of the AT\,2016dah photometry.} 
\begin{center}
\begin{tabular}{lllllllll}
\hline
\# & PANSTARRS ID & R.A.\ [J2000] & Dec.\ [J2000] & $u'$ [mag] & $g$ [mag] & $r$ [mag] & $i$ [mag] & $z$ [mag] \\
\hline
042 & 156100111892817585 & 00:44:45.427 & 40:05:20.808 & $17.199\pm0.008$ & $15.8041\pm0.0057$ & $15.3620\pm0.0019$ & $15.1845\pm0.0025$ & $15.1476\pm0.0126$ \\ 
043 & 156070111940992190 & 00:44:46.586 & 40:03:34.639 & \ldots & $18.3213\pm0.0041$ & $17.1903\pm0.0037$ & $16.6294\pm0.0033$ & $16.3634\pm0.0010$ \\ 
044 & 156140111960610721 & 00:44:47.060 & 40:07:00.256 & $18.373\pm0.013$ & $16.9415\pm0.0034$ & $16.5210\pm0.0066$ & $16.3430\pm0.0045$ & $16.3010\pm0.0065$ \\ 
045 & 156220111972291992 & 00:44:47.337 & 40:11:04.077 & \ldots & $18.9742\pm0.0050$ & $17.8405\pm0.0110$ & $17.3388\pm0.0050$ & $17.0969\pm0.0052$ \\ 
046 & 156270111997920146 & 00:44:47.954 & 40:13:28.612 & \ldots & $18.8844\pm0.0061$ & $18.5681\pm0.0078$ & $18.4496\pm0.0066$ & $18.4517\pm0.0103$ \\ 
047 & 156180112021090440 & 00:44:48.507 & 40:08:59.410 & \ldots & $17.2182\pm0.0031$ & $16.6114\pm0.0029$ & $16.3756\pm0.0031$ & $16.2984\pm0.0046$ \\ 
048 & 156170112036494555 & 00:44:48.883 & 40:08:41.704 & \ldots & $17.7909\pm0.0037$ & $17.3121\pm0.0048$ & $17.1169\pm0.0041$ & $17.0537\pm0.0062$ \\
\hline
\end{tabular}
\end{center}
\end{table}
\end{landscape}

\clearpage

\begin{landscape}
\begin{table}
\caption{SDSS sources used for calibration of the AT\,2017fyp photometry.}\label{17full_standards} 
\begin{center}
\begin{tabular}{lllllllll}
\hline
\# & SDSS ID & R.A.\ [J2000] & Dec.\ [J2000] & $u'$ [mag] & $g$ [mag] & $r$ [mag] & $i$ [mag] & $z$ [mag] \\
\hline
001 & J004502.95+395457.1 & 00:45:02.960 & 39:54:57.139 & $18.268\pm0.013$ & $17.159\pm0.005$ & $16.844\pm0.005$ & $16.723\pm0.005$ & $16.683\pm0.009$\\ 
002 & J004502.96+394717.8 & 00:45:02.967 & 39:47:17.826 & $18.498\pm0.019$ & $16.936\pm0.004$ & $16.380\pm0.004$ & $16.316\pm0.004$ & $16.192\pm0.008$\\ 
003 & J004508.22+394731.5 & 00:45:08.220 & 39:47:31.546 & $17.904\pm0.014$ & $16.350\pm0.003$ & $15.703\pm0.003$ & $15.582\pm0.003$ & $15.447\pm0.005$\\ 
004 & J004508.44+395313.0 & 00:45:08.444 & 39:53:13.038 & $18.243\pm0.017$ & $17.034\pm0.004$ & $16.598\pm0.004$ & $16.382\pm0.004$ & $16.359\pm0.008$\\ 
005 & J004508.44+395312.9 & 00:45:08.447 & 39:53:12.991 & $18.290\pm0.013$ & $17.051\pm0.005$ & $16.578\pm0.004$ & $16.408\pm0.005$ & $16.343\pm0.008$\\ 
006 & J004509.99+395522.6 & 00:45:09.999 & 39:55:22.681 & $18.902\pm0.026$ & $17.539\pm0.005$ & $17.034\pm0.004$ & $16.753\pm0.005$ & $16.671\pm0.011$\\ 
007 & J004510.00+395522.7 & 00:45:10.003 & 39:55:22.721 & $18.967\pm0.018$ & $17.559\pm0.005$ & $17.005\pm0.005$ & $16.795\pm0.005$ & $16.699\pm0.009$\\
\hline
\end{tabular}
\end{center}
\end{table}
\end{landscape}

\clearpage

\begin{table*}
\caption{Photometry of nova AT\,2016dah as obtained by ASAS-SN \citep[also see][]{2016ATel.9245....1N}, iPTF, and the LT.}\label{16full_photometry} 
\begin{center}
\begin{tabular}{lllllllll}
\hline
Date & $\Delta t$ & \multicolumn{2}{c}{MJD $57000+$} & Telescope and & Exposure & Filter & SNR & Photometry \\
\cline{3-4}
[UT] & [days] & Start & End & Instrument & [s] & & & [mag]\\
\hline
2016 Jul 16.067 & \phantom{0}\phantom{0}4.107 & 585.064 & 585.069 & LT IO:O & $3 \times 120$ & $u'$ & 119.0 & $15.908 \pm 0.048$ \\ 
2016 Jul 17.069 & \phantom{0}\phantom{0}5.109 & 586.067 & 586.072 & LT IO:O & $3 \times 120$ & $u'$ & 141.8 & $16.254 \pm 0.047$ \\ 
2016 Jul 18.059 & \phantom{0}\phantom{0}6.099 & 587.057 & 587.061 & LT IO:O & $3 \times 120$ & $u'$ & \phantom{0}68.5 & $16.630 \pm 0.048$ \\ 
2016 Jul 19.055 & \phantom{0}\phantom{0}7.095 & 588.052 & 588.057 & LT IO:O & $3 \times 120$ & $u'$ & \phantom{0}41.0 & $16.960 \pm 0.053$ \\ 
2016 Jul 22.095 & \phantom{0}10.135 & 591.093 & 591.097 & LT IO:O & $3 \times 120$ & $u'$ & \phantom{0}65.9 & $17.256 \pm 0.051$ \\ 
2016 Jul 25.050 & \phantom{0}13.090 & 594.047 & 594.052 & LT IO:O & $3 \times 120$ & $u'$ & \phantom{0}52.3 & $17.551 \pm 0.051$ \\
\hline
\end{tabular}
\end{center}
\end{table*}

\clearpage

\begin{table*}
\contcaption{Photometry of nova AT\,2016dah as obtained by ASAN-SN, iPTF, and the LT.}
\begin{center}
\begin{tabular}{lllllllll}
\hline
Date & $\Delta t$ & \multicolumn{2}{c}{MJD $57000+$} & Telescope and & Exposure & Filter & SNR & Photometry \\
\cline{3-4}
[UT] & [days] & Start & End & Instrument & [s] & & & [mag]\\
\hline
2016 Sep 06.100 & \phantom{-}56.140 & 637.098 & 637.102 & LT IO:O & $3 \times 120$ & $V$ & \phantom{0}26.4 & $21.646 \pm 0.043$ \\ 
2016 Sep 13.992 & \phantom{-}64.032 & 644.990 & 644.994 & LT IO:O & $3 \times 120$ & $V$ & \phantom{0}\phantom{-}8.2 & $21.946 \pm 0.133$ \\ 
2016 Sep 24.097 & \phantom{-}74.137 & 655.091 & 655.102 & LT IO:O & $3 \times 300$ & $V$ & \phantom{0}12.6 & $22.057 \pm 0.087$ \\ 
2016 Sep 28.115 & \phantom{-}78.155 & 659.109 & 659.120 & LT IO:O & $3 \times 300$ & $V$ & \phantom{0}22.1 & $22.140 \pm 0.051$ \\ 
2016 Dec 07.955 & 148.995 & 729.945 & 729.966 & LT IO:O & $3 \times 600$ & $V$ & \phantom{0}\phantom{-}2.6 & $(22.855 \pm 0.419)$ \\ 
\hline
2016 Aug 25.40 & & \ldots & \ldots & iPTF CFH12K &  & $g'$ & \ldots & $>20.5$  \\
\hline
\end{tabular}
\end{center}
\end{table*}

\clearpage

\begin{table*}
\contcaption{Photometry of nova AT\,2016dah as obtained by ASAN-SN, iPTF, and the LT.}
\begin{center}
\begin{tabular}{lllllllll}
\hline
Date & $\Delta t$ & \multicolumn{2}{c}{MJD $57000+$} & Telescope and & Exposure & Filter & SNR & Photometry \\
\cline{3-4}
[UT] & [days] & Start & End & Instrument & [s] & & & [mag]\\
\hline
2016 Aug 27.065 & \phantom{-}46.105 & 627.063 & 627.067 & LT IO:O & $3 \times 120$ & $r'$ & \phantom{0}74.7 & $20.791 \pm 0.015$ \\ 
2016 Aug 29.128 & \phantom{-}48.168 & 629.126 & 629.131 & LT IO:O & $3 \times 120$ & $r'$ & \phantom{0}58.1 & $20.836 \pm 0.019$ \\ 
2016 Sep 06.105 & \phantom{-}56.145 & 637.103 & 637.107 & LT IO:O & $3 \times 120$ & $r'$ & \phantom{0}46.4 & $21.342 \pm 0.023$ \\ 
2016 Sep 13.997 & \phantom{-}64.037 & 644.995 & 644.999 & LT IO:O & $3 \times 120$ & $r'$ & \phantom{0}11.3 & $21.726 \pm 0.096$ \\ 
2016 Sep 24.108 & \phantom{-}74.148 & 655.102 & 655.113 & LT IO:O & $3 \times 300$ & $r'$ & \phantom{0}24.5 & $21.966 \pm 0.044$ \\ 
2016 Sep 28.126 & \phantom{-}78.166 & 659.121 & 659.131 & LT IO:O & $3 \times 300$ & $r'$ & \phantom{0}26.2 & $22.182 \pm 0.042$ \\
\hline
\end{tabular}
\end{center}
\end{table*}

\begin{table*}
\caption{Photometry of nova AT\,2017fyp as obtained by the LT, LCO, {\it Gaia}, and ATLAS \citep[see][]{2017TNSTR.852....1T}.}\label{17full_photometry} 
\begin{center}
\begin{tabular}{lllllllll}
\hline
Date & $\Delta t$ & \multicolumn{2}{c}{MJD $57000+$} & Telescope and & Exposure & Filter & SNR & Photometry \\
\cline{3-4}
[UT] & [days] & Start & End & Instrument & [s] & & & [mag]\\
\hline
2017 Aug 10.115 & \phantom{0}\phantom{0}5.035 & \phantom{0}976.112 & \phantom{0}976.117 & LT IO:O & $3 \times 120$ & $u'$ & 189.2 & $17.250 \pm 0.010$ \\ 
2017 Aug 12.047 & \phantom{0}\phantom{0}6.967 & \phantom{0}978.045 & \phantom{0}978.049 & LT IO:O & $3 \times 120$ & $u'$ & 182.5 & $17.220 \pm 0.011$ \\ 
2017 Aug 14.106 & \phantom{0}\phantom{0}9.026 & \phantom{0}980.104 & \phantom{0}980.108 & LT IO:O & $3 \times 120$ & $u'$ & 242.6 & $17.101 \pm 0.009$ \\ 
2017 Aug 18.206 & \phantom{0}13.126 & \phantom{0}984.204 & \phantom{0}984.208 & LT IO:O & $3 \times 120$ & $u'$ & 205.1 & $17.678 \pm 0.009$ \\ 
2017 Aug 20.213 & \phantom{0}15.133 & \phantom{0}986.211 & \phantom{0}986.216 & LT IO:O & $3 \times 120$ & $u'$ & 142.5 & $17.894 \pm 0.012$ \\ 
2017 Aug 22.165 & \phantom{0}17.085 & \phantom{0}988.163 & \phantom{0}988.166 & LT IO:O & $2 \times 120$ & $u'$ & \phantom{0}13.6 & $18.286 \pm 0.082$ \\
\hline
\end{tabular}
\end{center}
\end{table*}

\begin{table*}
\contcaption{Photometry of nova AT\,2017fyp as obtained by the LT, LCO, {\it Gaia}, and ATLAS.} 
\begin{center}
\begin{tabular}{lllllllll}
\hline
Date & $\Delta t$ & \multicolumn{2}{c}{MJD $57000+$} & Telescope and & Exposure & Filter & SNR & Photometry \\
\cline{3-4}
[UT] & [days] & Start & End & Instrument & [s] & & & [mag]\\
\hline
2017 Aug 23.142 & \phantom{0}18.062 & \phantom{0}989.140 & \phantom{0}989.144 & LT IO:O & $3 \times 120$ & $B$ & 191.4 & $18.713 \pm 0.016$ \\ 
2017 Aug 24.569 & \phantom{0}19.489 & \phantom{0}990.567 & \phantom{0}990.571 & LT LCO Spectral & $3 \times 120$ & $B$ & \phantom{0}43.9 & $18.958 \pm 0.032$ \\ 
2017 Aug 25.065 & \phantom{0}19.985 & \phantom{0}991.063 & \phantom{0}991.068 & LT IO:O & $3 \times 120$ & $B$ & 157.3 & $18.994 \pm 0.018$ \\ 
2017 Aug 30.207 & \phantom{0}25.127 & \phantom{0}996.204 & \phantom{0}996.209 & LT IO:O & $3 \times 120$ & $B$ & 129.6 & $19.320 \pm 0.016$ \\ 
2017 Sep 04.974 & \phantom{0}30.894 & 1001.973 & 1001.975 & LT IO:O & $3 \times 60$ & $B$ & \phantom{0}28.5 & $19.588 \pm 0.041$ \\ 
2017 Sep 08.151 & \phantom{0}34.071 & 1005.150 & 1005.152 & LT IO:O & $3 \times 60$ & $B$ & \phantom{0}25.2 & $19.768 \pm 0.047$ \\
\hline
\end{tabular}
\end{center}
\end{table*}

\clearpage

\begin{table*}
\contcaption{Photometry of nova AT\,2017fyp as obtained by the LT, LCO, {\it Gaia}, and ATLAS.} 
\begin{center}
\begin{tabular}{lllllllll}
\hline
Date & $\Delta t$ & \multicolumn{2}{c}{MJD $57000+$} & Telescope and & Exposure & Filter & SNR & Photometry \\
\cline{3-4}
[UT] & [days] & Start & End & Instrument & [s] & & & [mag]\\
\hline
2017 Sep 23.082 & \phantom{0}49.002 & 1020.081 & 1020.083 & LT IO:O & $3 \times 60$ & $r'$ & \phantom{0}17.0 & $19.099 \pm 0.064$ \\ 
2017 Sep 27.992 & \phantom{0}53.912 & 1024.991 & 1024.994 & LT IO:O & $3 \times 60$ & $r'$ & \phantom{0}94.2 & $19.438 \pm 0.012$ \\ 
2017 Oct 07.945 & \phantom{0}63.865 & 1034.944 & 1034.946 & LT IO:O & $3 \times 60$ & $r'$ & \phantom{0}56.5 & $19.806 \pm 0.019$ \\ 
2017 Oct 23.061 & \phantom{0}78.981 & 1050.060 & 1050.062 & LT IO:O & $3 \times 60$ & $r'$ & \phantom{0}36.8 & $20.162 \pm 0.030$ \\ 
2017 Oct 28.090 & \phantom{0}84.010 & 1055.088 & 1055.091 & LT IO:O & $3 \times 60$ & $r'$ & \phantom{0}42.1 & $20.373 \pm 0.026$ \\ 
2017 Dec 21.929 & 138.849 & 1109.928 & 1109.931 & LT IO:O & $3 \times 60$ & $r'$ & --- & $>19.4$ \\
\hline
\end{tabular}
\end{center}
\end{table*}
\end{appendix}
\end{document}